\documentclass[floatfix,twocolumn,showpacs,preprintnumbers,amsmath,amssymb,pre,superscriptaddress]{revtex4-1}
\usepackage{color}
\usepackage[usenames,dvipsnames,svgnames,table]{xcolor}
\usepackage[colorlinks=true,linkcolor=blue,urlcolor=blue,citecolor=blue]{hyperref}

\usepackage{mathtools}
\usepackage{graphicx}
\usepackage{dcolumn}
\usepackage{array}
\usepackage{lipsum}
\usepackage{bm}
\usepackage{subfigure}
\usepackage{amssymb}
\usepackage{multirow}
\usepackage{tabularx}
\usepackage{amsmath}
\usepackage{braket}
\graphicspath{{plots/}}

\DeclareRobustCommand{\orderof}{\ensuremath{\mathcal{O}}}
\renewcommand{\vec}[1]{\mathbf{#1}}

\newcommand{\beq}{\begin{equation}}
\newcommand{\eeq}{\end{equation}}
\newcommand{\bea}{\begin{eqnarray}}
\newcommand{\eea}{\end{eqnarray}}

\begin{document}
\title{Dynamic properties of the warm dense electron gas: an \textit{ab initio}\\ path integral Monte Carlo approach }

\author{Paul Hamann}
\affiliation{Institut f\"ur Theoretische Physik und Astrophysik, Christian-Albrechts-Universit\"at zu Kiel,
 Leibnizstra{\ss}e 15, D-24098 Kiel, Germany}

\author{Tobias Dornheim}
\email{t.dornheim@hzdr.de}

\affiliation{Center for Advanced Systems Understanding (CASUS), G\"orlitz, Germany}


\author{Jan Vorberger}

\affiliation{Helmholtz-Zentrum Dresden-Rossendorf, Bautzner Landstra{\ss}e 400, D-01328 Dresden, Germany}

\author{Zhandos A.~Moldabekov}

\affiliation{Institute for Experimental and Theoretical Physics, Al-Farabi Kazakh National University, 71 Al-Farabi str.,  
  050040 Almaty, Kazakhstan}
  \affiliation{Institute of Applied Sciences and IT, 40-48 Shashkin Str., 050038 Almaty,Kazakhstan}

\author{Michael Bonitz}

\affiliation{Institut f\"ur Theoretische Physik und Astrophysik, Christian-Albrechts-Universit\"at zu Kiel,
 Leibnizstra{\ss}e 15, D-24098 Kiel, Germany}

\begin{abstract}
There is growing interest in warm dense matter (WDM) -- an exotic state on the border between condensed matter and plasmas. Due to the simultaneous importance of quantum and correlation effects WDM is complicated to treat theoretically. A key role has been played by \textit{ab initio} path integral Monte Carlo (PIMC) simulations, and recently extensive results for thermodynamic quantities have been obtained. The first extension of PIMC simulations to  the dynamic structure factor of the uniform electron gas were reported by Dornheim \textit{et al.} [Phys. Rev. Lett. \textbf{121}, 255001 (2018)]. This was based on an accurate reconstruction of the dynamic local field correction. Here we extend this concept to other dynamical quantities of the warm dense electron gas including the dynamic susceptibility, the dielectric function and the conductivity. \end{abstract}

\maketitle

\section{Introduction}

The uniform electron gas (UEG) is one of the most important model systems in quantum physics and theoretical chemistry~\cite{quantum_theory,loos,review}. Despite its apparent simplicity, it offers a wealth of interesting effects like collective excitations (plasmons)~\cite{pines} and Wigner crystallization at low density~\cite{wigner,drummond_wigner}. At zero temperature, most static properties of the UEG have been known for decades from ground-state quantum Monte Carlo (QMC) simulations~\cite{gs2,moroni2,spink,ortiz1,ortiz2}, and the accurate parametrization of these results~\cite{vwn,perdew,perdew_wang,gori-giorgi1,gori-giorgi2} has been pivotal for the spectacular success of density functional theory (DFT) regarding the description of real materials~\cite{dft_review,burke_perspective}.

The recent interest in warm dense matter (WDM)---an extreme state that occurs, e.g., in astrophysical objects~\cite{militzer1,saumon1,becker} and on the pathway towards inertial confinement fusion~\cite{hu_ICF}---has made it necessary to extend these considerations to finite temperatures. More specifically, WDM is defined by two characteristic parameters, which are both of the order of unity: a) the density parameter (Wigner-Seitz radius) $r_s=\overline{r}/a_\textnormal{B}$ (with $\overline{r}$ and $a_\textnormal{B}$ being the average particle distance and first Bohr radius) and b) the degeneracy temperature $\theta=k_\textnormal{B}T/E_\textnormal{F}$ (with $E_\textnormal{F}$ being the usual Fermi energy~\cite{torben_eur}). Moreover, WDM is nowadays routinely realized in the laboratory, see Ref.~\cite{falk_wdm} for a review on experimental techniques, and many important results have been achieved over the last years \cite{Fletcher2015,exp1,exp2,exp3,exp4}.

On the other hand, the theoretical description of WDM is most challenging~\cite{wdm_book,new_POP} due to the complicated interplay of 1) thermal excitations, 2) quantum degeneracy effects, and 3) Coulomb scattering. For example, the non-negligible coupling strength rules out perturbation expansions~\cite{kas1,kas3}. Semi-classical approaches like molecular dynamics using quantum potentials~\cite{MD1,MD2} fail due to strong quantum degeneracy effects and exchange effects. While \textit{ab initio} QMC methods are, in principle, capable to take into account all of these effects exactly simultaneously, they are afflicted with the notorious fermion sign problem (FSP, see Ref.~\cite{dornheim_sign_problem} for an accessible topical introduction). In particular, the FSP leads to an exponential increase in computation time both upon increasing the system size and decreasing the temperature~\cite{dornheim_sign_problem,loh,troyer}, and has been shown to be $NP$-hard for a particular class of Hamiltonians~\cite{troyer}.

For this reason, it took more than three decades after the celebrated ground-state results for the UEG by Ceperley and Alder~\cite{gs2} to obtain accurate data in the WDM regime~\cite{brown_ethan,schoof_prl,dornheim_prl,groth_prl,malone2,dornheim_pop}. This was achieved by developing and combining new QMC methods that are available in complementary parameter regions~\cite{dornheim,dornheim2,groth,dornheim3,dornheim_cpp,blunt,malone1,malone2,dornheim_neu}.
These efforts have culminated in the first accurate parametrizations of the exchange--correlation (XC) free energy $f_\textnormal{xc}$ of the UEG~\cite{groth_prl,ksdt,karasiev_status}, which provide a complete description of the UEG over the entire WDM regime. Moreover, these results allow for thermal DFT simulations~\cite{mermin_dft,rajagopal_dft} in the local density approximation, and recent studies \cite{kushal,karasiev_importance} have revealed that thermal XC effects are indeed crucial to correctly describe aspects of WDM such as microscopic density fluctuations and the behaviour of hydrogen bonds at finite temperature.

While being an important milestone, it is clear that a more rigorous theory of WDM requires to go beyond the local density approximation. In this context, the key information is given by the response of the UEG to a time-dependent external perturbation, which is fully characterized by the dynamic density response function~\cite{kugler1}
\begin{eqnarray}\label{eq:chi}
\chi(q,\omega) = \frac{\chi_0(q,\omega)}{1-\tilde v(q)\left(1-G(q,\omega)\right)\chi_0(q,\omega) } \quad .
\end{eqnarray}
Here $\tilde v(q)=4\pi/q^2$ is the Fourier transform of the Coulomb potential, $\chi_0(q,\omega)$ denotes the density response function of an ideal (i.e., non interacting) Fermi gas, and $G(q,\omega)$ is commonly known as the dynamic local field correction (LFC).
More specifically, setting $G(q,\omega)=0$ in Eq.~(\ref{eq:chi}) leads to a mean-field description of the density response (known as random phase approximation, RPA), and, consequently, $G(q,\omega)$ contains the full frequency- and wave-number-resolved description of XC effects.

Obviously, such information is vital for many applications. This includes the construction of advanced, non-local XC-functions for DFT simulations~\cite{burke_ac,lu_ac,patrick_ac,goerling_ac}, and the exchange--correlation kernel for the time-dependent DFT (TDDFT) formalism~\cite{without_chihara}. Moreover, we mention the incorporation of XC-effects into quantum hydrodynamics~\cite{new_POP,diaw1,diaw2,zhanods_hydro}, the construction of effective ion-ion potentials~\cite{ceperley_potential,zhandos1,zhandos2}, and the interpretation of WDM experiments~\cite{siegfried_review,kraus_xrts}. Finally, the dynamic density response of the UEG can be directly used to compute many material properties such as the electronic stopping power~\cite{stopping2,stopping}, electrical and thermal conductivities~\cite{Desjarlais:2017,Veysman:2016}, and energy transfer rates~\cite{jan_relax}.

Yet, obtaining accurate data for $\chi(q,\omega)$ and related quantities has turned out to be very difficult. In the ground state, Moroni \textit{et al.}~\cite{moroni2,moroni} obtained QMC data for the density response function and LFC in the static limit (i.e., for $\omega=0$) by simulating a harmonically perturbed system and subsequently measuring the actual response. Remarkably, this computationally expensive strategy is not necessary at finite temperatures, as the full wave-number dependence of the static limit of the density response $\chi(q)=\chi(q,0)$ can be obtained from a single simulation of the unperturbed system~\cite{dynamic_folgepaper,dornheim_ML}, see Eq.~(\ref{eq:static_chi}) below. In this way, Dornheim \textit{et al.}~\cite{dornheim_ML} were recently able to provide extensive path integral Monte Carlo (PIMC) data both for $\chi(q)$ and $G(q)$ for the warm dense UEG, which, in combination with the ground-state data~\cite{moroni2,cdop}, has allowed to construct a highly accurate machine-learning based representation of $G(q;r_s,\theta)$ covering the entire relevant WDM regime. Moreover, PIMC results for the static density response have been presented for the strongly coupled electron liquid regime ($r_s\geq20$) \cite{dornheim_EL}, and the high-energy density limit ($r_s\leq0.5$) \cite{dornheim_HED}.
Finally, we mention that even the nonlinear regime has been studied by the same group~\cite{dornheim_nonlinear}.

The last unexplored dimension is then the frequency-dependence of $\chi(q,\omega)$, which constitutes a formidable challenge that had remained unsolved even at zero temperature. Since time-dependent QMC simulations suffer from an additional dynamical sign problem~\cite{dynamic_sign_problem,dynamic_sign_problem2}, previous results for the dynamic properties of the UEG were based on perturbation theories like the nonequilibrium Green function formalism at finite temperature~\cite{kwong,kas1} or many-body theory in the ground state~\cite{takada1,takada2}.

Fortunately, PIMC simulations give direct access to the intermediate scattering function $F$ [defined in Eq.~(\ref{eq:define_F})], but evaluated at imaginary times $\tau\in[0,\beta]$, which is related to the dynamic structure factor $S(q,\omega)$ by a Laplace transform,
\begin{eqnarray}\label{eq:S_F}
F(q,\tau) = \int_{-\infty}^\infty \textnormal{d}\omega\ S(q,\omega) e^{-\tau\omega} \quad .
\end{eqnarray}
The numerical solution of Eq.~(\ref{eq:S_F}) for $S(q,\omega)$ is a well-known, but notoriously difficult problem~\cite{jarrell}. While different approaches based on, e.g., Bayes theorem~\cite{mem_revisited} or genetic optimization~\cite{gift,gift2} exist, it was found necessary to include additional information into this reconstruction procedure to sufficiently constrain the results for $S(q,\omega)$. In order to do so, we have introduced a stochastic sampling procedure~\cite{dornheim_dynamic,dynamic_folgepaper} for the dynamic LFC, which allows to automatically fulfill a number of additional exact properties. Thus, we were able to present the first unbiased results for the dynamic structure factor of the warm dense UEG without any approximation regarding XC effects.

In the present work, we further extend these considerations and adapt our reconstruction procedure to obtain other dynamic properties of the UEG such as the dielectric function $\epsilon(q,\omega)$, the conductivity $\sigma(q,\omega)$ and the density response function $\chi(q,\omega)$ itself. 
Further, we analyze the respective accuracy of different quantities and find that the comparably large uncertainty in the dynamic LFC $G(q,\omega)$ has only small impact on physical properties like $\chi(q,\omega)$, $\epsilon(q,\omega)$, and $S(q,\omega)$, which are well constrained by the PIMC results. Thus, this work constitutes a proof-of-concept investigation and opens up new avenues for WDM theory, electron liquid theory and beyond.

The paper is organized as follows: In Sec.~\ref{sec:theory} we summarize the main formulas of linear response theory and introduce our PIMC approach to the dynamic local field correction. In Sec.~\ref{sec:results} we present our \textit{ab initio} simulation results for the local field correction, the dynamic structure factor, the density response function, the dielectric function, and the dynamic conductivity.
We conclude with a summary and outlook in Sec.~\ref{sec:summary} where we give a concise list of future extensions of our work.

\section{Theory and simulation idea\label{sec:theory}}

\subsection{Path integral Monte Carlo}

The basic idea of the path integral Monte Carlo method~\cite{imada,cep} is to evaluate thermodynamic expectation values by stochastically sampling the density matrix,
\begin{eqnarray}\label{eq:rho}
\rho(\mathbf{R}_a,\mathbf{R}_b,\beta) = \bra{\mathbf{R}_a} e^{-\beta\hat H} \ket{\mathbf{R}_b}\ ,
\end{eqnarray}
in coordinate space, with $\beta=1/k_\textnormal{B}T$ being the inverse temperature and $\mathbf{R}=(\mathbf{r}_1,\dots,\mathbf{r}_N)^T$ containing the coordinates of all $N$ particles.
Unfortunately, a direct evaluation of Eq.~(\ref{eq:rho}) is not possible since the kinetic and potential contributions $\hat K$ and $\hat V$ to the Hamiltonian do not commute,
\begin{eqnarray}\label{eq:convergence}
e^{-\beta\hat H} = e^{-\beta\hat K}e^{-\beta\hat V} + \mathcal{O}\left(
\beta^{2}\right)\ .
\end{eqnarray}
To overcome this issue, one typically employs a Trotter decomposition~\cite{trotter} and finally ends up with an expression for the (canonical) partition function $Z$ of the form
\begin{eqnarray}\label{eq:weight}
Z = \int \textnormal{d}\mathbf{X}\ W(\mathbf{X})\ ,
\end{eqnarray}
with the meta-variable $\mathbf{X}=(\mathbf{R}_0,\dots,\mathbf{R}_{P-1})^T$ being a so-called \textit{configuration}, which is taken into account according to the corresponding configuration weight $W(\mathbf{X})$.

\begin{figure}
\includegraphics[width=0.4147\textwidth]{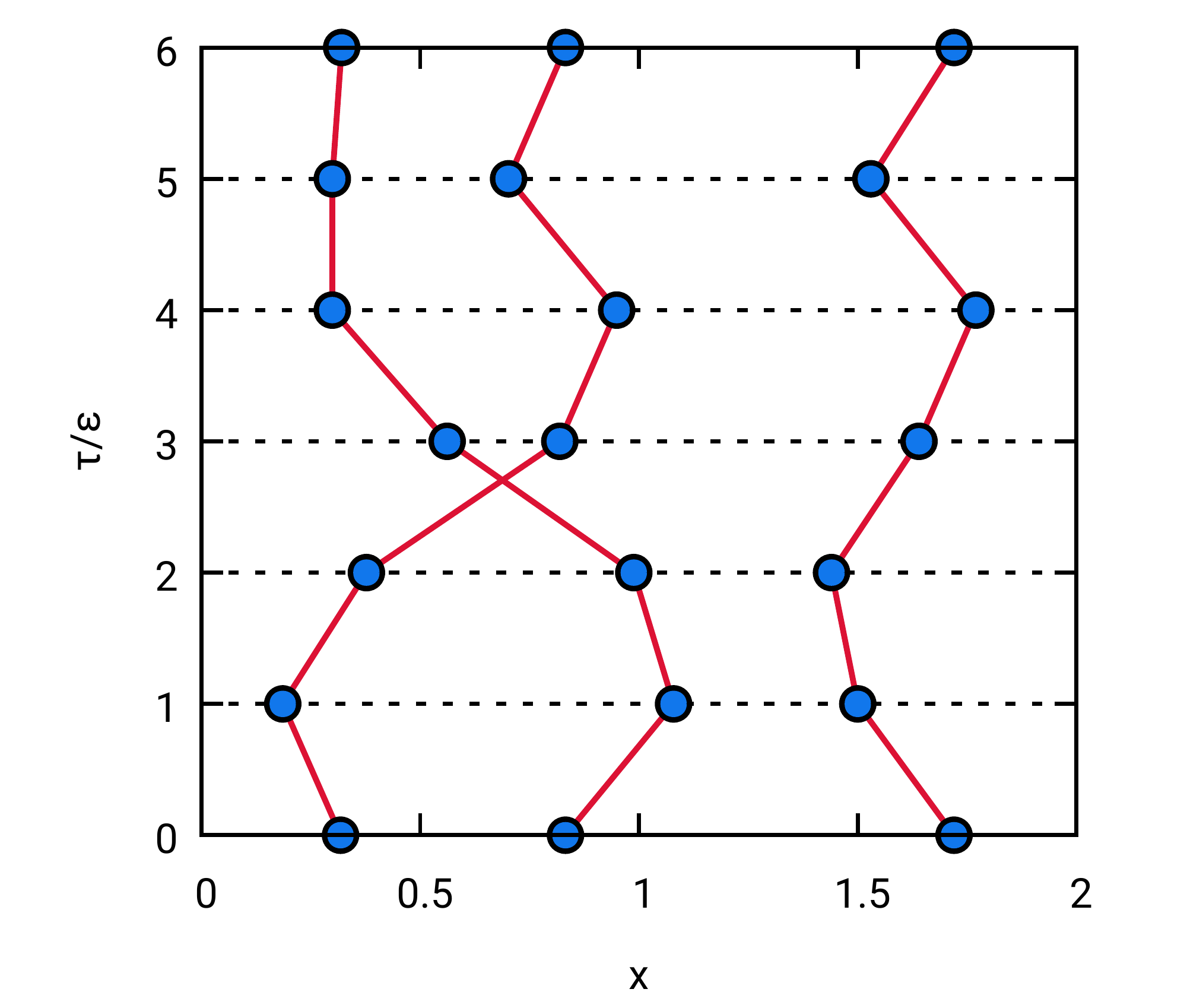}
\caption{\label{fig:PIMC}
Schematic illustration of Path Integral Monte Carlo---Shown is a configuration of $N=3$ electrons with $P=6$ imaginary--time propagators in the $x$-$\tau$ plane. Due to the single pair-exchange, the corresponding configuration weight $W(\mathbf{X})$ [cf.~Eq.~(\ref{eq:weight})] is negative. Reprinted from T.~Dornheim \textit{et al.}, \textit{J.~Chem.~Phys.}~\textbf{151}, 014108 (2019) \cite{dornheim_permutation_cycles} with the permission of AIP Publishing.
}
\end{figure}

This is illustrated in Fig.~\ref{fig:PIMC}, where we show an exemplary configuration of $N=3$ electrons. First and foremost, we note that each particle is now represented by an entire \textit{path} of $P=6$ coordinates located on different \textit{imaginary-time} slices, which are separated by a time-step of $\epsilon=\beta/P$. In particular, $P$ constitutes a convergence parameter within the PIMC method and has to be chosen sufficiently high to ensure unbiased results, cf.~Eq.~(\ref{eq:convergence}). For completeness, we mention that the convergence with respect to $P$ can, in principle, be accelerated by using higher-order factorizations of the density matrix, e.g., Refs.~\cite{HO1,HO2}. This, however, is not advisable for the present study, since it would also limit the number of $\tau$-points, on which the density--density correlation function $F(q,\tau)$ can be evaluated, cf.~Eq.~(\ref{eq:define_F}) below.
In addition, the formulation of the PIMC method in imaginary-time allows for a straightforward evaluation of imaginary-time correlation functions, such as 
\begin{eqnarray}\label{eq:define_F}
F(q,\tau) = \frac{1}{N}  \braket{\hat\rho(q,\tau)\hat\rho(-q,0)}\,,
\end{eqnarray}
where the two density operators are simply evaluated at two different time slices.

For the PIMC method, one uses the Metropolis algorithm~\cite{metropolis} to stochastically generate a Markov chain of configurations $\mathbf{X}$, which can then be used to compute the thermodynamic expectation value of an arbitrary observable $\hat A$,
\begin{eqnarray}\label{eq:MC}
\braket{\hat A}_\textnormal{MC} = \frac{1}{N_\textnormal{MC}}\sum_{k=1}^{N_\textnormal{MC}} A\left(\mathbf{X}_i\right) \ .
\end{eqnarray}
Here $\braket{\hat A}_\textnormal{MC}$ denotes the Monte Carlo estimate, which converges towards the exact expectation value in the limit of a large number of random configurations
\begin{eqnarray}
\braket{\hat A} = \lim_{N_{MC}\to\infty} \braket{\hat A}_\textnormal{MC}\ ,
\end{eqnarray}
and the statistical uncertainty (error bar) decreases as
\begin{eqnarray}\label{eq:MC_error}
\Delta A_\textnormal{MC} \sim \frac{1}{\sqrt{N_\textnormal{MC}}}\ .
\end{eqnarray}
Thus, both the factorization error with respect to $P$ and the MC error [Eq.~(\ref{eq:MC_error})] can be made arbitrarily small, and the PIMC approach is \textit{quasi-exact}.

An additional obstacle is given by the fermion sign problem, which follows from the sign changes in the configuration weight $W(\mathbf{X})$ due to different permutations of particle paths. In particular, configurations with an odd number of pair permutations (such an example is shown in Fig.~\ref{fig:PIMC}) result in negative weights, which is a direct consequence of the antisymmetry of the fermionic density matrix under particle exchange. This leads to an exponential increase in computation time with increasing the system size $N$ or decreasing the temperature.
However, a more extensive discussion of the sign problem is beyond the scope of the present work, and has been presented elsewhere~\cite{dornheim_permutation_cycles,dornheim_sign_problem}.

For completeness, we mention that all PIMC data presented in this work have been obtained using a canonical adaption~\cite{mezza} of the worm algorithm introduced by Boninsegni \textit{et al.}~\cite{boninsegni1,boninsegni2}.

\subsection{Stochastic sampling of the dynamic LFC}\label{ss:sampling}

In this section, we describe the numerical solution of Eq.~(\ref{eq:S_F}) based on the stochastic sampling of the dynamic local field correction $G(q,\omega)$ introduced in Refs.~\cite{dornheim_dynamic,dynamic_folgepaper}. In principle, the task at hand is to find a trial solution $S_\textnormal{trial}(q,\omega)$, which, when being inserted into Eq.~(\ref{eq:S_F}), reproduces the PIMC data for $F(q,\tau)$ within the given Monte Carlo error bars. This, however, is a notoriously difficult and, in fact, ill-posed problem~\cite{jarrell}, as different trial solutions with distinct features might reproduce $F_\textnormal{PIMC}(q,\tau)$ within the given confidence interval.

To further constrain the space of possible trial solutions, one might consider the frequency moments of the dynamic structure factor, which are defined as
\begin{eqnarray}\label{eq:moment}
\braket{\omega^k} = \int_{-\infty}^\infty \textnormal{d}\omega\ \omega^k \,S(q,\omega)  \, .
\end{eqnarray}
For the UEG, four frequency moments are known from different sum-rules, namely $k=-1,0,1,3$. The corresponding equations are summarized in Ref.~\cite{dynamic_folgepaper}, and need not be repeated here.

For some applications~\cite{dynamic_alex1,dynamic_alex2}, the frequency moments have been shown to significantly improve the quality of the reconstruction procedure. For the UEG, on the other hand, the combined information from $F(q,\tau)$ and $\braket{\omega^k}$ is not sufficient to fully determine the shape and position of the plasmon peaks.

To overcome this issue, Dornheim and co-workers~\cite{dornheim_dynamic,dynamic_folgepaper} proposed to further constrain the space of possible solutions by automatically fulfilling a number of exact properties of the dynamic LFC $G(q,\omega)$.
The central equation for this strategy is the well-known fluctuation--dissipation theorem~\cite{quantum_theory}, which gives a relation between $S(q,\omega)$ and the dynamic density response function $\chi(q,\omega)$,
\begin{eqnarray}\label{eq:FDT}
S(\mathbf{q},\omega) = - \frac{ \textnormal{Im}\chi(\mathbf{q},\omega)  }{ \pi n (1-e^{-\beta\omega})}\ .
\end{eqnarray}
The latter is then expressed in terms of the density response function of the noninteracting system and the dynamic LFC, see Eq.~(\ref{eq:chi}) above. Therefore, using Eqs.~(\ref{eq:FDT}) and (\ref{eq:chi}), we have recast the reconstruction problem posed by Eq.~(\ref{eq:S_F}) into the search for a suitable trial solution for the dynamic LFC, $G_\textnormal{trial}(q,\omega)$.

The important point is that many additional exact properties of $G(q,\omega)$ are known from theory. Since, again, all formulas are listed in Ref.~\cite{dynamic_folgepaper}, here we give a brief summary:
\begin{enumerate}
    \item The Kramers-Kronig relations give a direct connection between the real and imaginary parts of $G(q,\omega)$ in the form of a frequency-integral.
    \item Re $G(q,\omega)$ and Im $G(q,\omega)$ are even and odd functions with respect to $\omega$, respectively.
    \item The imaginary part vanishes for $\omega=0$ and $\omega\to\infty$.
    \item The static ($\omega=0$) limit of Re $G(q,\omega)$ can be directly obtained from
    \begin{eqnarray}\label{eq:static_chi}
\chi(\mathbf{q},0) = -n\int_0^\beta \textnormal{d}\tau\ F(\mathbf{q},\tau)\ ,
\end{eqnarray}
and Eq.~(\ref{eq:chi}). Further, an accurate neural-net representation of $G(q,\omega=0)$ was presented in Ref.~\cite{dornheim_ML}.
    \item The high-frequency ($\omega\to\infty$) limit of Re $G(q,\omega)$ can be computed from the static structure factor $S(q)$ and the exchange--correlation contribution to the kinetic energy. The latter is obtained from the accurate parametrization of the exchange--correlation free energy $f_\textnormal{xc}$ by Groth \textit{et al.}~\cite{groth_prl}.
\end{enumerate}

To generate trial solutions for $G(q,\omega)$ that automatically fulfill these constraints, we follow an idea by Dabrowski~\cite{dabrowski} and introduce an extended  Pad\'e formula of the form
\begin{eqnarray}\label{eq:parametrization}
\textnormal{Im}\,G(\mathbf{q},\omega) = \frac{ a_0\omega + a_1\omega^3 + a_2\omega^5 }{ \left( b_0 + b_1\omega^2 \right)^c }\ ,
\end{eqnarray}
where $a_{0-2}$, $b_{0,1}$, and $c$ are chosen randomly. The corresponding real part of this trial solution $G_\textnormal{trial}(q,\omega)$ is subsequently computed from the Kramers-Kronig relation, see Ref.~\cite{dynamic_folgepaper} for a more detailed discussion.

During the reconstruction procedure, we 1) randomly generate a large set of parameters $\textbf{T}=\{a_i,b_i,c\}$, 2) use these to obtain both Im $G_\textnormal{trial}(q,\omega)$ and Re $G_\textnormal{trial}(q,\omega)$, 3) Insert these into Eq.~(\ref{eq:chi}) to compute the corresponding $\chi_\textnormal{trial}(q,\omega)$, 4) insert the latter into the fluctuation--dissipation theorem to get a dynamic structure factor $S_\textnormal{trial}(q,\omega)$, and 5) compare $S_\textnormal{trial}(q,\omega)$ to our PIMC data for $F(q,\tau)$ (for all $\tau\in[0,\beta]$) and the frequency moments $\braket{\omega^k}$.
The small subset of $M$ trial parameters $\mathbf{T}$ that reproduce both $F(q,\tau)$ and $\braket{\omega^k}$ are kept to obtain the final result for physical quantities of interest, like $S(q,\omega)$ itself, but also $\chi(q,\omega)$ and $\epsilon(q,\omega)$.

For example, the final solution for the dynamic structure factor is given by
\begin{eqnarray}\label{eq:final_average}
S_\textnormal{final}(\mathbf{q},\omega) = \frac{1}{M} \sum_{i=1}^M S_{\textnormal{trial},i}(\mathbf{q},\omega)\ .
\end{eqnarray}
Moreover, this approach allows for a straightforward estimation of the associated uncertainty as the corresponding variance
\begin{eqnarray}\label{eq:delta_S}
\Delta S(\mathbf{q},\omega) = \left( 
\frac{1}{M}\sum_{i=1}^M\left[ 
S_{\textnormal{trial},i}(\mathbf{q},\omega) \right.\right. \\  - \left.\left. S_\textnormal{final}(\mathbf{q},\omega)
\right]^2
\right)^{1/2} \nonumber\ .
\end{eqnarray}

\subsection{Dielectric function and inverse dielectric function}\label{ss:epsilon}
Having obtained \textit{ab initio} results for the density response function $\chi(q,\omega)$, it is straightforward to obtain the dynamic retarded dielectric function~\cite{bonitz_book} as well as the inverse dielectric function,
\begin{align}
    \epsilon(q,\omega) &= 1 - \tilde v(q)\Pi(q,\omega),\\
    \epsilon^{-1}(q,\omega) &= 1 + \tilde v(q)\chi(q,\omega)\,,
    \label{eq:epsm1}
\end{align}
where $\Pi$ is the retarded polarization function. Its relation to the density response function is
\begin{align}
    \Pi(q,\omega) = \epsilon(q,\omega) \chi(q,\omega) = \frac{\chi(q,\omega)}{1+\tilde v(q)\chi(q,\omega)}\,.
    \label{eq:pi-chi}
\end{align}
In the limiting case of an ideal Fermi gas, $\chi \to \chi_0$, where $\chi_0$ is the Lindhard response function (but at finite temperature), and Eq.~(\ref{eq:pi-chi}) yields the RPA polarization, and the dielectric function becomes
\begin{align}
    \epsilon^{\rm RPA}(q,\omega) &= 1 - \tilde v(q)\chi_0(q,\omega). 
    \label{eq:eps_rpa}
\end{align}

Correlation effects, i.e. deviations from $\chi_0$, can be expressed in terms of the dynamic local field correction $G(q,\omega)$, and the dielectric function can be written as
\begin{equation}
    \epsilon(q,\omega) = 1 - \frac{\tilde v(q)\chi_0(q,\omega)}{1+\tilde v(q) G(q,\omega) \chi_0(q,\omega)}. \label{eq:eps-G}
\end{equation}
While $\epsilon^{-1}$ is commonly used in linear response theory, $\epsilon$ emerges naturally in electrodynamics, e.g.~Ref.~\cite{alexandrov_book}, and it is of prime importance for the description of plasma oscillations.

Let us summarize a few definitions and important properties  of the retarded dielectric function. 
\begin{enumerate}
    \item Since $\chi(q,\omega)$ describes a causal response, $\epsilon^{-1}(q,\omega)$ is an analytic function in the upper frequency half-plane. Real and imaginary parts are connected via the Kramers-Kronig relation for real frequencies.
    \item If $G(q,\omega)$ is computed via an \textit{ab initio} QMC procedure \cite{dornheim_dynamic}, also the dielectric function has \textit{ab initio} quality. We will call this result $\epsilon^{\rm DLFC}(q,\omega)$.
    \item Another important approximation is obtained by replacing, in Eq.~(\ref{eq:eps-G}), $G(q,\omega) \to G(q,0)=G(q)$. This is still a dynamic dielectric function which will be denoted by $\epsilon^{\rm SLFC}(q,\omega)$. Comparison to the full dynamic treatment revealed that this static approximation provides an accurate description of the dynamic structure factor for $r_s\lesssim 4$, for all wave numbers~\cite{dornheim_dynamic}.
    \item Since the static limit of the response function is real and negative:
    \begin{equation}
        \chi(q,0) \leq 0,
    \end{equation} 
    which is a necessary prerequisite for the stability of any system, it immediately follows for the static dielectric function that:
    \begin{equation}
        \epsilon(q,0)^{-1} < 1,
    \end{equation}
    which implies $\epsilon(q,0) \not\in (0,1]$.
    \item The static long wavelength limit is real and related to the compressibility $K$,
    \begin{equation}
        K^{-1} = n^2 \frac{\partial \mu}{\partial n}
    \end{equation}
    via
    \begin{equation}\label{eq:compressibility}
      \lim\limits_{q\rightarrow 0}\epsilon(q,0) = 1 + \tilde v(q) n^2 K\,,
  \end{equation} 
  where $n$ is the density and $\mu$ the chemical potential \cite{quantum_theory}. The practical evaluation of the compressibility is given by Eq.~(\ref{eq:real_CSR}) below.

\end{enumerate}

\subsection{Dynamic conductivity}\label{ss:conductivity}

Having the dielectric function at hand, it is straightforward to compute further dynamic linear response quantities. An example is the dynamical conductivity $\sigma(q,\omega)$, that follows from the response of the current density to an electric field with the result
\begin{align}
\epsilon(q,\omega) = 1 + \frac{4\pi i}{\omega}\sigma(q,\omega)\,.
\label{eq:conductivity}
\end{align}
This can be transformed into an expression for the conductivity in terms of the RPA response function and the dynamic local field correction, using Eq.~(\ref{eq:eps-G}),
\begin{align}
\sigma(q,\omega) & = i \frac{\omega}{4\pi}\frac{\tilde v(q)\chi_0(q,\omega)}{1+ \tilde v(q)G(q,\omega)\chi_0(q,\omega)}\,
    \label{eq:sigma-G}
\\
&= i \frac{\omega}{4\pi}\tilde v(q) \Pi(q,\omega)\,,
\nonumber
\end{align}
where $\Pi$ is the longitudinal polarization function (\ref{eq:pi-chi}).\\

The analytical properties of the  polarization function in RPA at finite temperature were thoroughly investigated in many papers, including Refs.~\cite{Deutsch_1978, arista-brandt_84, Dandrea_86}. Following these works, it is straightforward to find various limiting cases for the conductivity in RPA which are valuable for comparison to the correlated results presented in Sec.~\ref{ss:other}.

1.) At $q\ll q_F$ (and for arbitrary frequency), we have for the real part of the conductivity in RPA:

\begin{equation}
\frac{{\rm Re}~\sigma(q,\omega)}{\omega_p}=\frac{2\sqrt{3}}{9\pi^2}r_s^{3/2} \frac{\left(\frac{q}{q_F}\right)^{-3}\left(\frac{\omega}{\omega_p}\right)^2}{1+\exp\left[\frac{\alpha r_s}{\theta}\frac{(\omega/\omega_p)^2}{(q/q_F)^2}-\eta\right]} ,
\label{eq:RPA_limit2}
\end{equation}
where $\alpha=3 (4/9\pi)^{4/3}\simeq 0.221$ and $\eta=\beta \mu$.

2.) At $\omega\ll q v_F$ (i.e., $\omega/\omega_p\ll (q/q_F)\times 2.13/\sqrt{r_s}$) and arbitrary wavenumber, the real part of the conductivity in RPA reads
\begin{equation}
\frac{{\rm Re}~\sigma(q,\omega)}{\omega_p}=\frac{2\sqrt{3}}{9\pi^2}r_s^{3/2} \frac{\left(\frac{q}{q_F}\right)^{-3}\left(\frac{\omega}{\omega_p}\right)^2}{1+\exp\left[\frac{(q/q_F)^2}{4\theta}-\eta\right]} .
\label{eq:RPA_limit1}
\end{equation}

3.) For the imaginary part of the conductivity in RPA, at high frequencies,  $\omega\gg \hbar q^2/(2m)$ and $\omega\gg qv_F$, we find the following result:
\begin{eqnarray}
 \label{eq:RPA_limit3}
  \frac{{\rm Im}~\sigma(q,\omega)}{\omega_p}=&&  \frac{1}{4\pi}\left(\frac{\omega_p}{\omega}\right)\\\nonumber
  &+& \left[\frac{3}{16 ~r_s}\left(\frac{9\pi}{4}\right)^{1/3}\frac{\langle v^2\rangle}{v_F^2}\left(\frac{q}{q_F}\right)^2 \right.\\\nonumber
    &+&\left.
  \frac{3}{64 ~r_s}\left(\frac{9\pi}{4}\right)^{1/3}\left(\frac{q}{q_F}\right)^4\right]\left(\frac{\omega_p}{\omega}\right)^3\\\nonumber
 &+& \frac{9\pi}{64 ~r_s^2}\left(\frac{9\pi}{4}\right)^{2/3}\frac{\langle v^4\rangle}{v_F^4}\left(\frac{q}{q_F}\right)^4\left(\frac{\omega_p}{\omega}\right)^5+...,
\end{eqnarray}
where $\langle ...\rangle$ indicates an average with the finite temperature Fermi function. Analytical parametrizations for the moments, $\frac{\langle v^2\rangle}{v_F^2}$ and $\frac{\langle v^4\rangle}{v_F^4}$, are given in the Appendix.

If one neglects terms of the order $\orderof\left( (\omega_p/\omega)^{\,3}\right)$ and higher, i.e. retains only the $\orderof\left( (\omega_p/\omega)^{\,1}\right)$ order term, the often used high frequency limit for the RPA conductivity is recovered (e.g., in the Drude conductivity)\cite{Reinholz_2000},   
\begin{equation}
  \frac{{\rm Im}~\sigma(q,\omega)}{\omega_p}=  \frac{1}{4\pi}\left(\frac{\omega_p}{\omega}\right).
 \label{eq:RPA_limit4}
\end{equation}

\section{Numerical Results\label{sec:results}}

\subsection{Density correlation function}\label{ss:f-results}

\begin{figure}
    \centering
   \hspace*{-0.5cm} \includegraphics[width=1.15\linewidth]{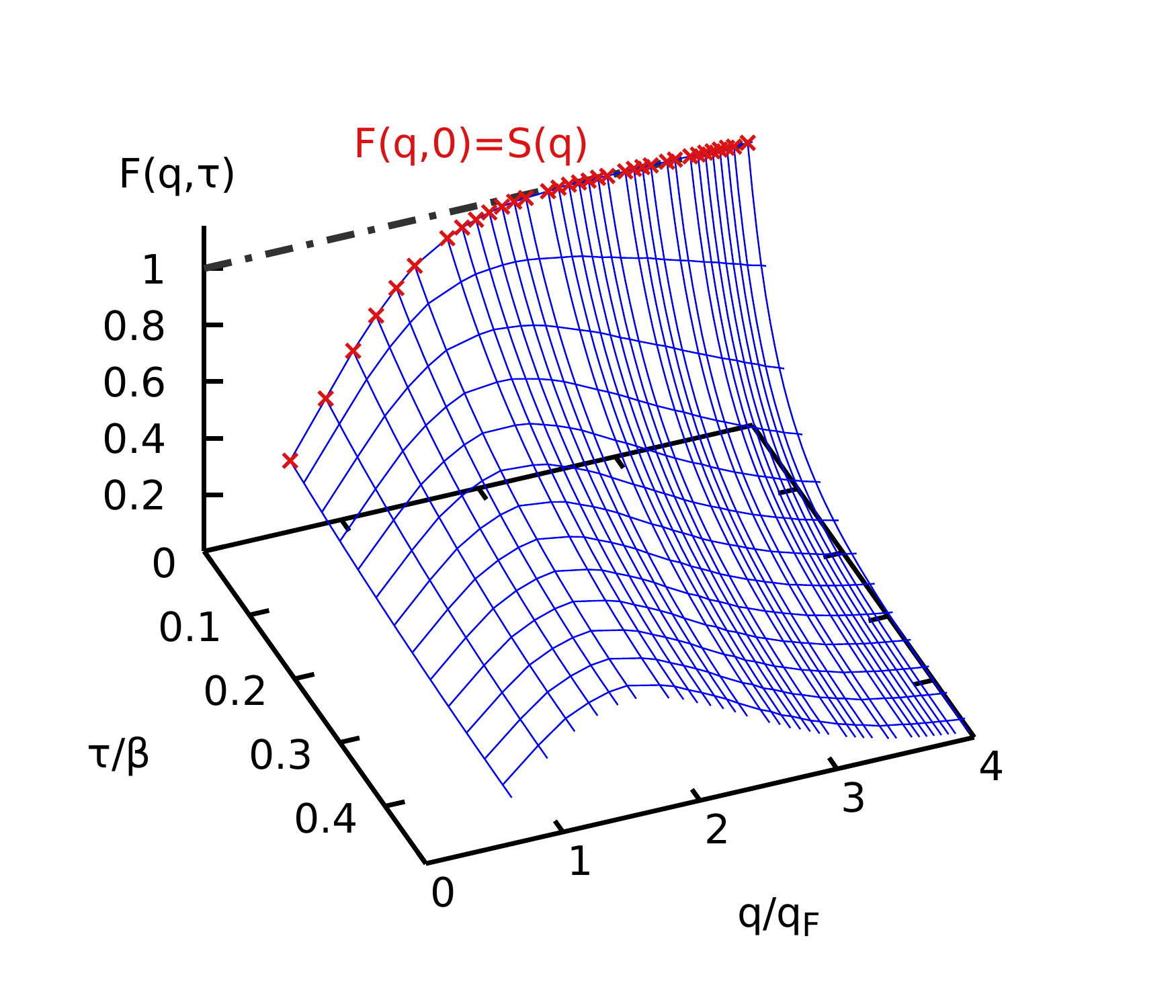}
    \caption{PIMC data for the imaginary-time density--density correlation function $F(q,\tau)$ for $N=34$ unpolarized electrons at $r_s=4$ and $\theta=1$ (WDM conditions).  The red crosses correspond to the static structure factor $S(q)=F(q,0)$.}
    \label{fig:FPLOT}
\end{figure}

Let us begin the investigation of our numerical results with a brief discussion of the imaginary-time density--density correlation function $F(q,\tau)$, which constitutes the most important input for the reconstruction procedure. To this end, we show $F(q,\tau)$ in the $q$-$\tau$-plane for $r_s=4$ and $\theta=1$ in Fig.~\ref{fig:FPLOT}. Since a physically meaningful interpretation of this quantity is rather difficult, here we restrict ourselves to a summary of some basic properties. First and foremost, we note that $F$ approaches the static structure factor (see the red crosses in Fig.~\ref{fig:FPLOT}) in the limit of small $\tau$, $F(q,0)=S(q)$. Moreover, $F$ is symmetric in the imaginary time around $\tau=\beta/2$ [i.e., $F(q,\tau)=F(q,\beta-\tau)$], and it is thus fully sufficient to show only the range of $\tau\in[0,\beta/2]$.

Regarding physical parameters, the density--temperature combination depicted in Fig.~\ref{fig:FPLOT} corresponds to a metallic density in the WDM regime. This is somewhat reflected in Fig.~\ref{fig:FPLOT} by the amount of structure in the surface plot, in particular the maximum around $q\approx1.5q_\textnormal{F}$. For example, for larger coupling strength the UEG forms an electron liquid and $F(q,\tau)$ exhibits a more pronounced structure with several maxima and minima~\cite{dornheim_EL}.
For decreasing $r_s$, on the other hand, electronic correlation effects become less important and one approaches the high-energy-density regime, where $F(q,\tau)$ exhibits even less structure than for the present example, and the only maximum is shifted to smaller values of $q/q_\textnormal{F}$, see Ref.~\cite{dornheim_HED}.

\begin{figure}
    \centering
    \includegraphics[width=\linewidth]{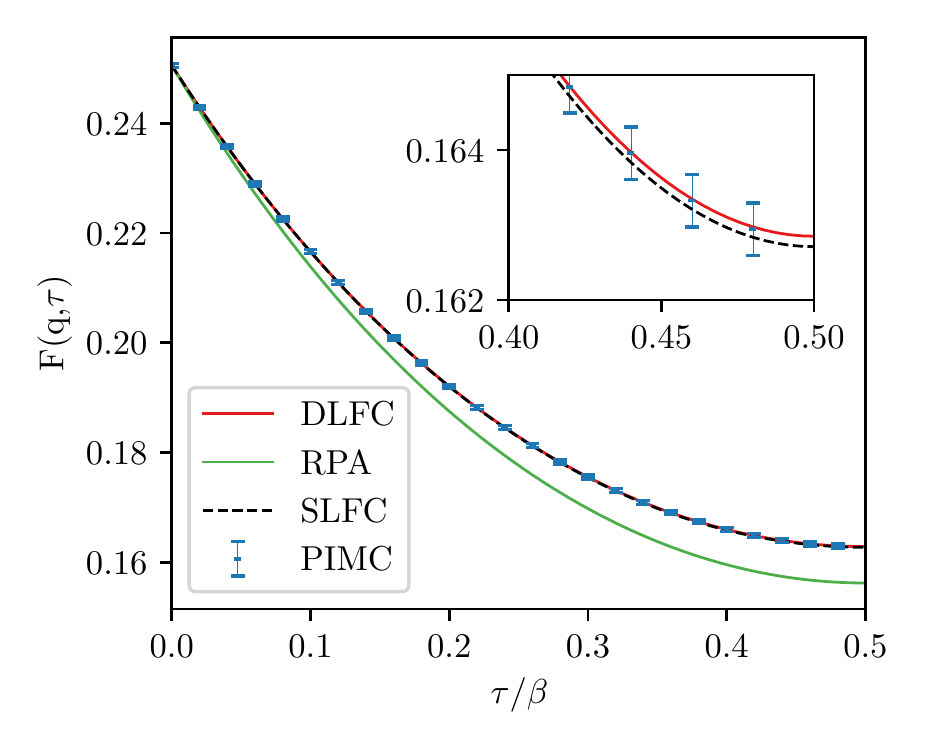}
    \caption{Imaginary-time density-density correlation function $F(q,\tau)$ for $r_s=4$, $\theta=1$ at $q/q_F\approx0.63$. The PIMC data is compared to the reconstruction result.}
    \label{fig:F_imag}
\end{figure}

\begin{figure*}
\centering
\includegraphics[width=0.88\textwidth]{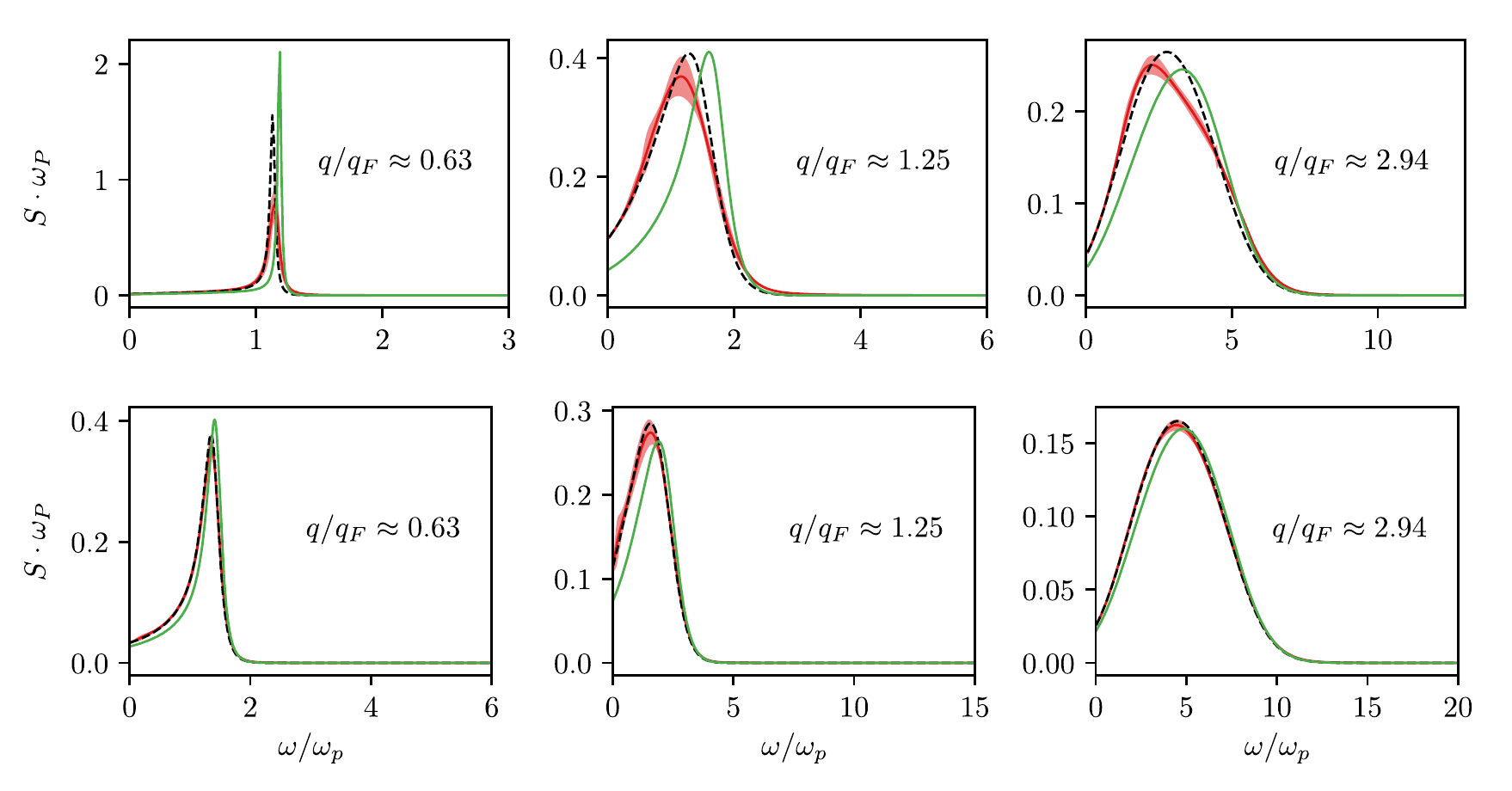}
\caption{Dynamic structure factor $S(q,\omega)$ at $\theta=1$ for three wave numbers. Top: $r_s=10$, bottom: $r_s=4$. Green: RPA, dashed: SLFC, red: DLFC.}
\label{fig:dynSF}
\end{figure*}

Let us next briefly touch upon the utility of the imaginary-time density--density correlation function for the reconstruction of dynamic quantities like the dynamic structure factor $S(q,\omega)$.
This is illustrated in Fig.~\ref{fig:F_imag}, where we show the $\tau$-dependence of $F$ for a fixed wave number $q\approx0.63q_\textnormal{F}$.
The blue points correspond to our PIMC data, and the three curves have been obtained by inserting different solutions for the dynamic structure factor into Eq.~(\ref{eq:S_F}), see the bottom left panel of Fig.~\ref{fig:dynSF} for the corresponding depiction of $S(q,\omega)$.  Let us start with the green curve, which shows the random phase approximation (RPA). Evidently, the mean field description exhibits severe deviations from the exact PIMC data and is too low by $\sim10\%$ over the entire $\tau$-range. Moreover, this shift is not constant, and the RPA curve exhibits a faster decay with $\tau$ compared to the blue points. This is consistent with previous studies~\cite{dornheim_prl,review,dornheim_cpp} of the static structure factor $S(q)$, where RPA has been shown to give systematically too low results for all wave numbers.

In contrast, the dashed black curve has been obtained on the basis of the \emph{static approximation}, i.e., by setting $G(q,\omega)=G(q,0)$ in Eq.~(\ref{eq:chi}).
Evidently, this leads to a substantially improved imaginary-time density--density correlation function, and the black curve is within the Monte-Carlo error bars over the entire $\tau$-range. Finally, the solid red curve has been obtained by stochastically sampling the full frequency-dependence of $G(q,\omega)$ as described in Sec.~\ref{ss:sampling}. While this does lead to an even better agreement to the PIMC data, one cannot decide between the two solutions on the basis of $F(q,\tau)$ alone. This further illustrates the need for the incorporation of the exact constraints on the stochastic sampling of $G(q,\omega)$, as the static approximation and full frequency-dependence of the LFC lead to substantially different dynamic structure factors, but similar $F(q,\tau)$.

\subsection{Dynamic structure factor}\label{ss:s-results}

The next quantity of interest is the dynamic structure factor $S(q,\omega)$ itself, which is shown in Fig.~\ref{fig:dynSF} at the Fermi temperature for two different densities and for three different wave numbers each. In this context, we recall that $S(q,\omega)$ constitutes a key quantity for the full reconstruction of any dynamical property, as it is used as a measure of quality of the dynamic LFC $G(q,\omega)$ in the stochastic sampling procedure, see Sec.~\ref{ss:sampling}. Moreover, it is directly accessible in XRTS experiments~\cite{siegfried_review} and is of paramount importance for plasma diagnostics~\cite{kraus_xrts}, like the determination of the electronic temperature. For this reason, $S(q,\omega)$ has been extensively investigated in previous studies~\cite{dornheim_dynamic,dynamic_folgepaper,dornheim_FSC}. Very recently, Dornheim and Vorberger~\cite{dornheim_FSC} have found that the \textit{ab initio} PIMC results for $S(q,\omega)$ at the Fermi temperature are not afflicted with any significant finite-size error even for as few as $N=14$ unpolarized electrons. In addition, Dornheim \textit{et al.}~\cite{dornheim_dynamic} have presented results going from the WDM regime to the strongly coupled electron liquid regime, where $S(q,\omega)$ exhibits a negative dispersion relation (estimated from the maximum in the DSF), 
which might indicate the onset of an incipient excitonic mode~\cite{takada1,takada2,higuchi}.
For this reason, here we restrict ourselves to a brief discussion of the most important features.

The top row of Fig.~\ref{fig:dynSF} shows results for $r_s=10$ and the left, center, and right panels corresponds to $q\approx0.63q_\textnormal{F}$, $q\approx1.25q_\textnormal{F}$, and $q\approx2.94q_\textnormal{F}$, respectively. For the smallest wave number, we find a relatively sharp peak slightly above the plasma frequency, which can be identified as the plasmon of the UEG. Yet, while it is well known that this collective excitation is correctly described on the mean-field level (i.e., within RPA, green curve) for $q\to0$, we find substantial deviations between the three depicted data sets for $q\approx0.63q_\textnormal{F}$. More specifically, the static approximation (dashed black)
leads to a \emph{red-shift} compared to RPA, and a correlation-induced broadening. In addition, this trend becomes even more pronounced for the full reconstructed solution (red curve), which results in an almost equal position of the peak as the static solution, but is much broader.

The center panel corresponds to an intermediate wave number, and we find an even more pronounced red-shift. Remarkably, the static approximation performs very well and can hardly be distinguished from the full dynamic solution within the given confidence interval (red shaded area). Finally, the right panel shows the DSF for approximately thrice the Fermi wave number, where we also observe some interesting behaviour. First and foremost, all three spectra exhibit a similar width, which is much broader compared to the previous cases, as it is expected. Furthermore, including a local field correction in Eq.~(\ref{eq:chi}) leads to a red-shift, which is most pronounced for the red curve. Yet, the full frequency dependence of $G(q,\omega)$ leads to a nontrivial shape in $S(q,\omega)$, which is not captured by the black curve and leads to a substantially more pronounced negative dispersion relation compared to the static approximation~\cite{dornheim_dynamic}.

The bottom row of Fig.~\ref{fig:dynSF} corresponds to $r_s=4$, which is a metallic density in the WDM regime. The most striking difference to the electron liquid example above is that the spectra are comparably much broader for the same value of $q/q_\textnormal{F}$, which is a direct consequence of the increased density. Moreover, all three curves are in relatively good agreement, we observe only a small red-shift compared to RPA, and the static approximation fully describes the DSF. 
The same holds for the two larger wave numbers shown in the center and right panels, although here the red-shift to the mean-field description is somewhat larger.

In accordance with Ref.~\cite{dornheim_dynamic}, we thus conclude that the static approximation provides a nearly exact description of $S(q,\omega)$ for weak to moderate coupling ($r_s\lesssim5$) and constitutes a significant improvement over RPA even at the margins of the strongly coupled electron liquid regime.
The verification of this finding for other dynamic quantities like the dynamic density response function $\chi(q,\omega)$ [Sec.~\ref{ss:results-chi}] and dielectric function $\epsilon(q,\omega)$ [Sec.~\ref{ss:results-epsilon}] is one of the central goals of this work.

\subsection{Local field correction}\label{ss:results-dlfc}

\begin{figure}
    \centering
    \includegraphics[width=0.93\linewidth]{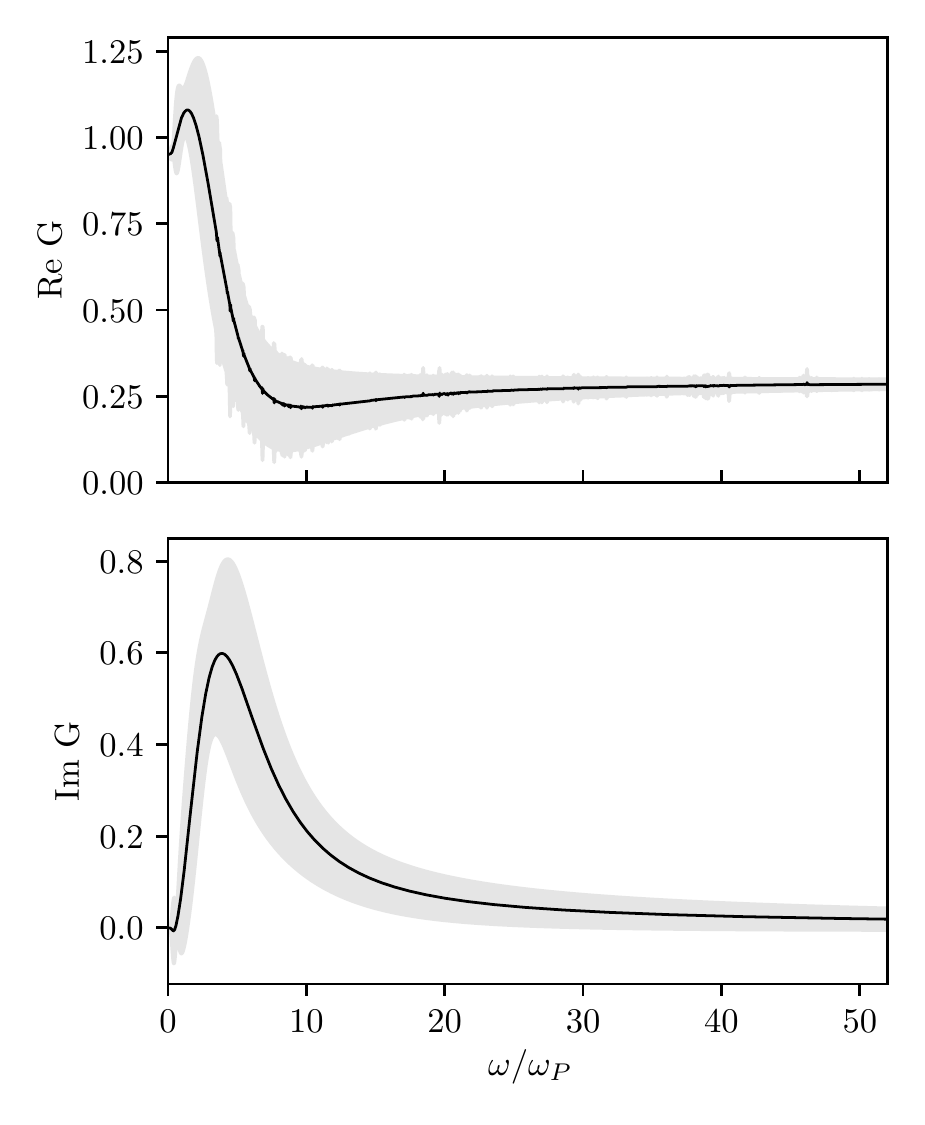}
    \caption{Average solution obtained for the real and imaginary parts of the dynamic local field correction, $G(q,\omega)$, at $\theta=1$, $r_s=6$, and $q\approx1.88\,q_F$. The shaded area corresponds to the range of valid reconstructions.}
    \label{fig:reg-img}
\end{figure}

Before we move on to the investigation of $\chi(q,\omega)$ and $\epsilon(q,\omega)$, it is well worth to briefly examine the reconstructed dynamic local field correction. In particular, the dynamic LFC constitutes the basis for the computation of all other dynamic properties from the \textit{ab initio} PIMC data and, thus, is of central importance for our reconstruction scheme. Since the stochastic sampling and subsequent elimination/verification of trial solutions for $G(q,\omega)$ has been extensively discussed by Groth \textit{et al.}~\cite{dynamic_folgepaper}, here we restrict ourselves to the discussion of a typical example shown in Fig.~\ref{fig:reg-img} for $\theta=1$, $r_s=6$, and $q\approx1.88q_\textnormal{F}$. These parameters are located at the margins of the WDM regime with a comparably large impact of electronic correlation effects and can be realized experimentally in hydrogen jets~\cite{Zastrau} and evaporation experiments, e.g. at the Sandia Z-machine~\cite{benage,karasiev_importance,low_density1,low_density2}.
Furthermore, the selected wave number is located in the most interesting regime, where the position of the maximum of $S(q,\omega)$ exhibits a non-monotonous behaviour and the impact of $G(q,\omega)$ is expected to be most pronounced.

In the top panel, we show the frequency-dependence of the real part of $G$, which exhibits a fairly nontrivial progression: starting from the exact static limit $G(q,0)$ directly known from our PIMC data for $\chi(q)$, the dynamic LFC exhibits a maximum around $\omega\approx1.4\omega_\textnormal{p}$ followed by shallow minimum around $\omega\sim10\omega_\textnormal{p}$ and then monotonically converges to the also exactly known $\omega\to\infty$ limit from below. Moreover, the associated uncertainty interval (light grey shaded area) is relatively small, and the maximum appears to be significant, whereas the minimum is probably not. At this point, we mention that the dynamic LFC is afflicted with the largest relative uncertainty of all reconstructed quantities considered in this work, which can be understood as follows: by design, the LFC contains the full information about exchange--correlation effects beyond RPA. In the limit of small and large wave numbers, the RPA is already exact and, consequently, the dynamic LFC has no impact on the reconstructed solutions for $S(q,\omega)$ that are compared to the PIMC data using Eq.~(\ref{eq:S_F}). Naturally, the same also holds for increasing temperature and density where the importance of $G(q,\omega)$ also vanishes~\cite{dornheim_HED}.

\begin{figure*}
    \centering
    \includegraphics[width=0.8\textwidth]{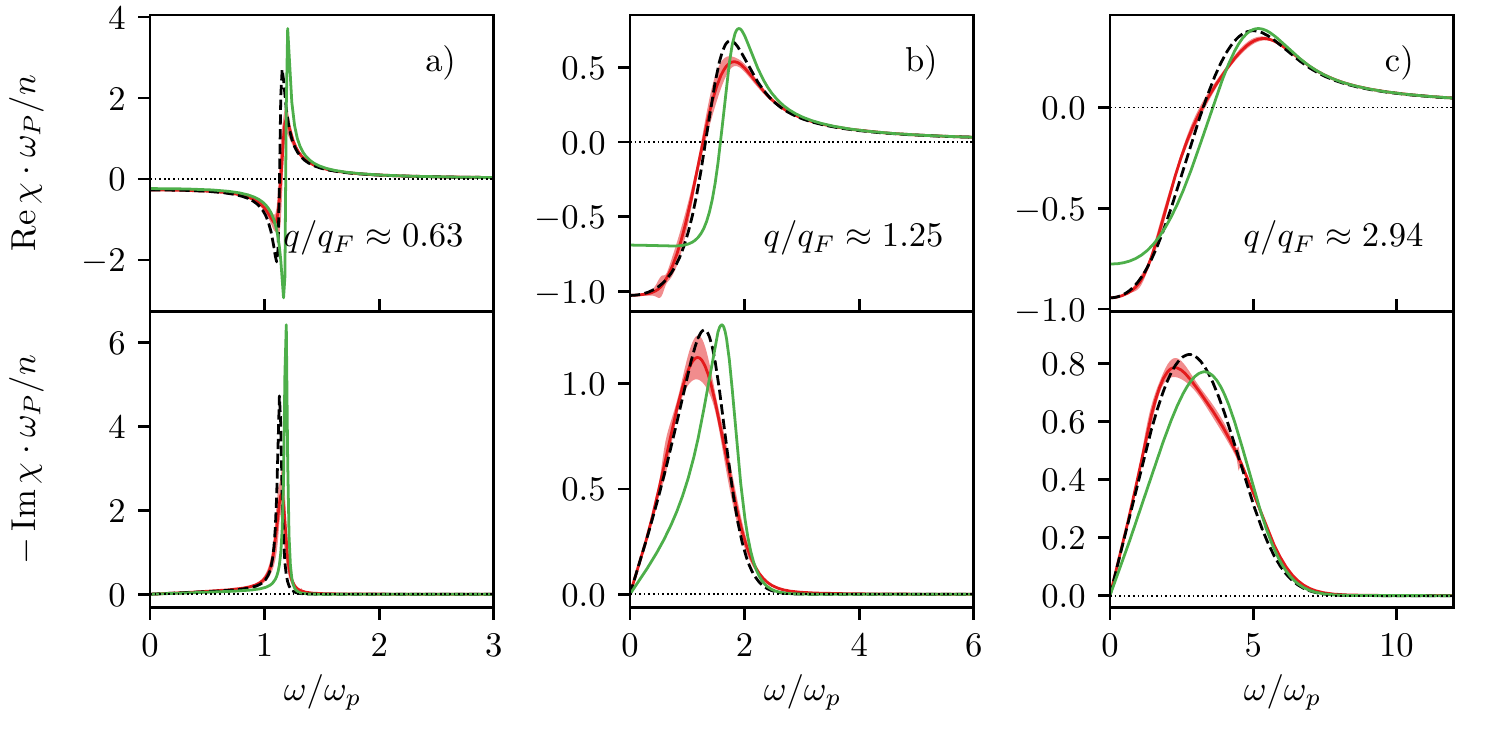}
    \includegraphics[width=0.8\textwidth]{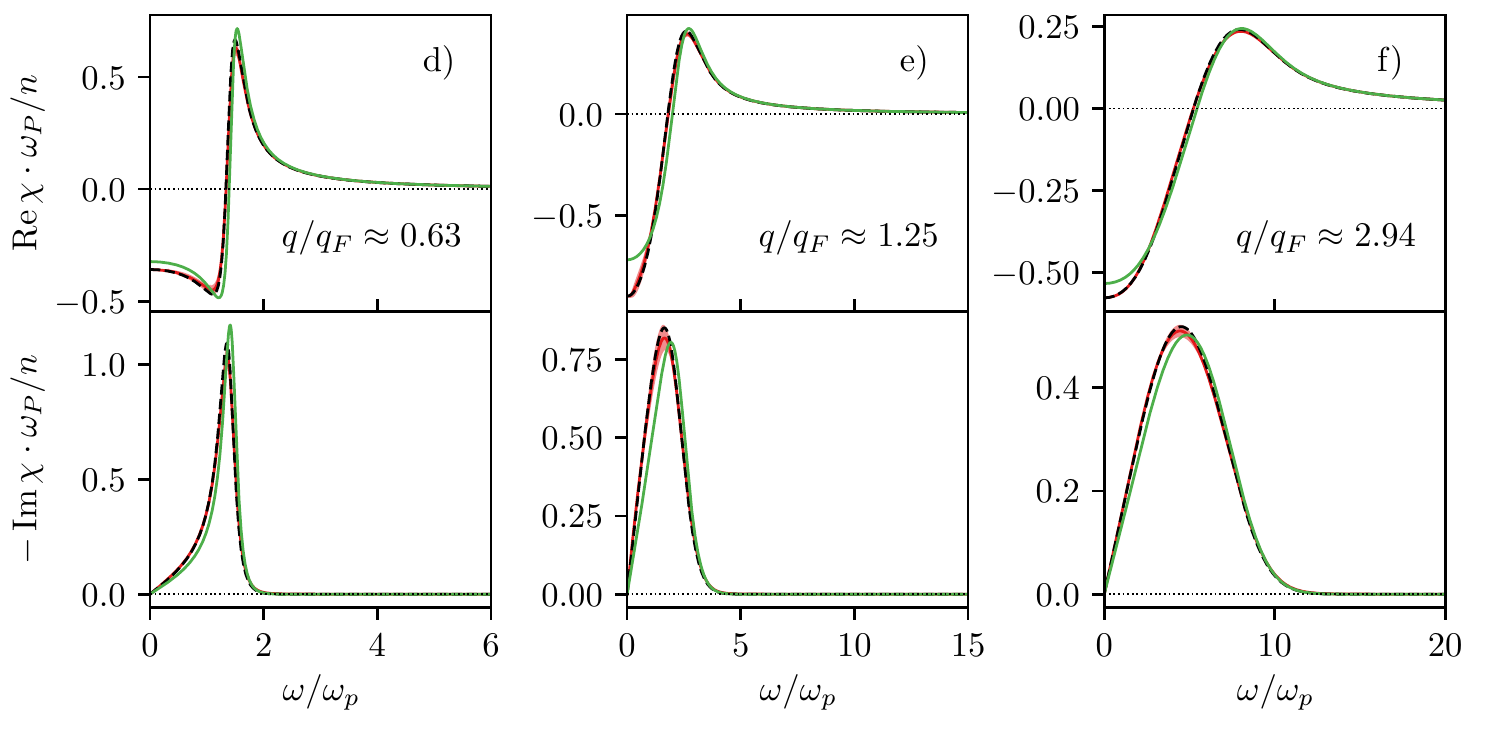}
    \caption{Dynamic density response function at $\theta=1$, for three wave numbers. a)--c): $r_s=10$; d)--f): $r_s=4$. Top (bottom) of each figure:  real (minus imaginary part). Comparison of RPA (green), and PIMC simulation results using the static (SLFC, dashed) and dynamic (DLFC, red) local field correction.}
    \label{fig:chi-rs10}
\end{figure*}

In the bottom panel of Fig.~\ref{fig:reg-img}, we show the corresponding imaginary part of $G$ for the same conditions. Interestingly, Im$G$ exhibits a seemingly less complicated behaviour featuring a single maximum around $\omega\approx5\omega_\textnormal{p}$, and vanishing both, in the high and low frequency limits from above.

We thus conclude that our reconstruction scheme allows us to obtain accurate results for the dynamic LFC when it has impact on physical observables like $S(q,\omega)$, i.e., precisely when it is needed in the first place. This allows for the intriguing possibility to construct a both $q$- and $\omega$-dependent representation of $G$ for some parameters, which could then be used for many applications like a real time-dependent DFT simulation without the adiabatic approximation.

\subsection{Dynamic density response function $\chi$}\label{ss:results-chi}
The present procedure to reconstruct the dynamic structure factor via a reconstruction of $G(q,\omega)$ can be straightforwardly extended to other dynamic quantities. The first example is the density response function defined by Eq.~(\ref{eq:chi}). Here we 
extend our preliminary results \cite{new_POP} and present more extensive data for two densities of interest.

\begin{figure*}
    \centering
    \includegraphics[width=0.4\linewidth]{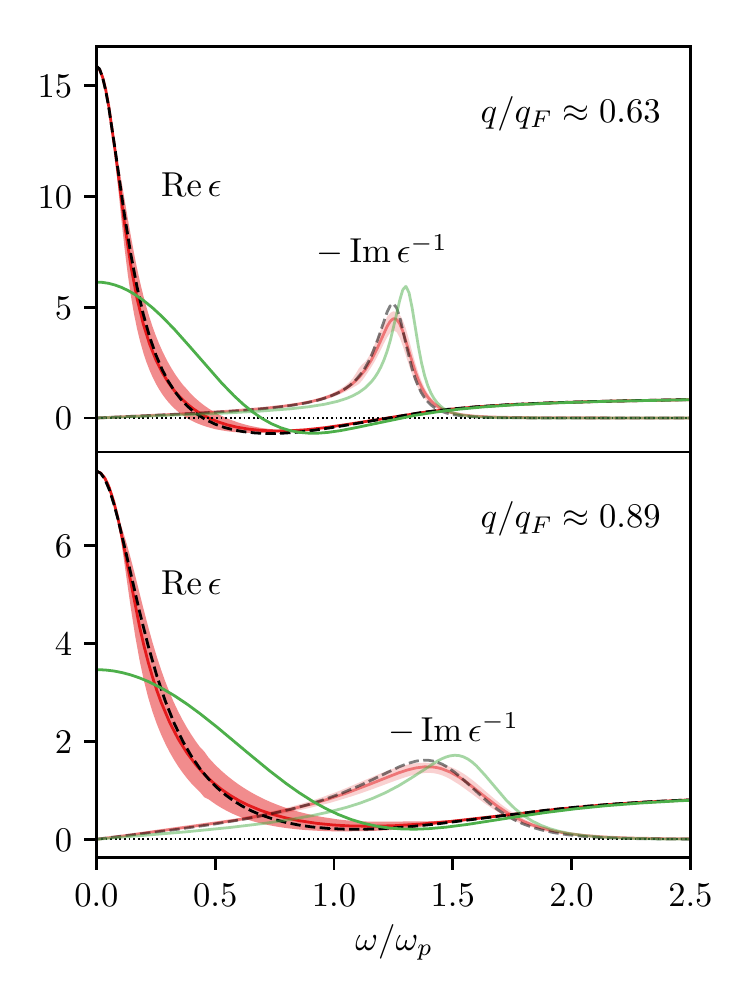}
    \includegraphics[width=0.4\linewidth]{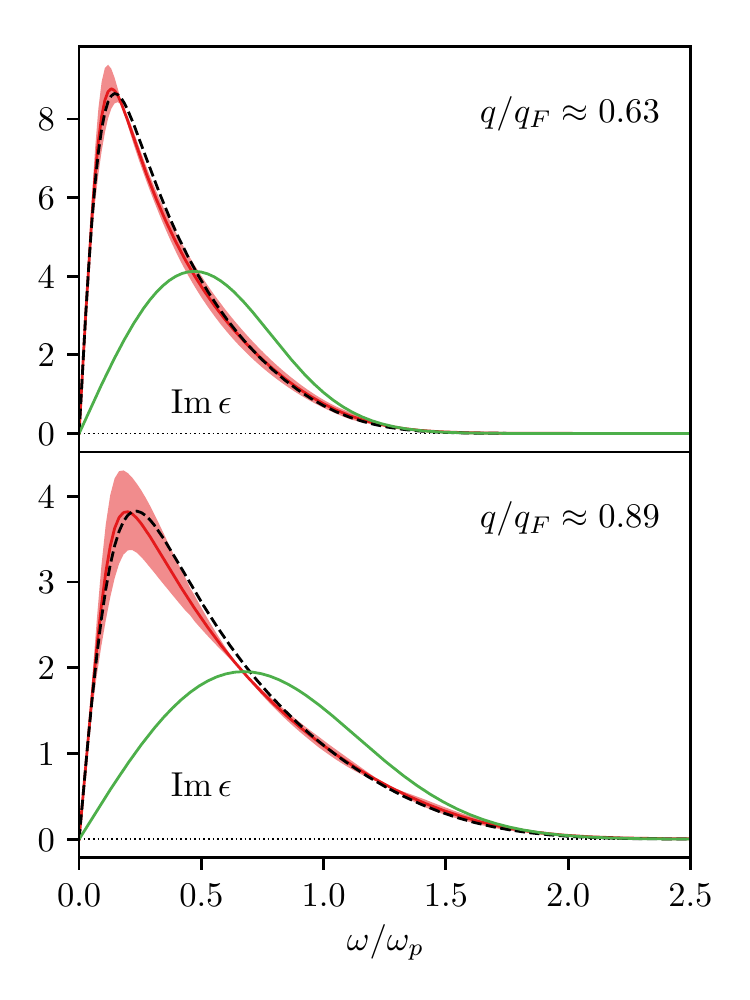}
    \caption{\textbf{Left}: Real part of the dielectric function $\epsilon(\vec{q},\omega)$ and imaginary part of the inverse dielectric function. \textbf{Right}: Imaginary part of the dielectric function, for $r_s=6$, $\theta=1$. 
    The peak of -Im$\epsilon^{-1}$ [and of $S(\vec{q},\omega)$] is in the vicinity of the second root of $\operatorname{Re}\epsilon$ (if roots exist, as in the upper figure). Green lines: RPA, red (dashed) lines: dynamic (static) PIMC results, for details see text. 
    }
   \label{fig:eps-rs6}
\end{figure*}

The top half of Fig.~\ref{fig:chi-rs10} corresponds to $r_s=10$ (strong coupling), and panels a)-c) to three interesting wave numbers. Similarly to the dynamic LFC discussed in the previous section, we find that the real part of $\chi$ exhibits a more complicated behaviour compared to the imaginary part. More specifically, the latter vanishes both in the high- and low-frequency limits, whereas the former attains a finite value in the static case. In addition, Re$\chi(q,\omega)$ has a remarkable structure in between,
with a pole-like structure owing to the Kramers-Kronig relation between imaginary and real part. This is the excitation of density fluctuations visible as a peak in the imaginary part, as translated to the real part and this structure.
 For the smallest depicted wave number, the excitation range is narrow and the position of the zero crossing of Re$\chi(q,\omega)$ almost exactly coincides with the position of the maximum in both Im$\chi(q,\omega)$ and $S(q,\omega)$.
 In contrast, a broader excitation at larger values of $q$ leads to a shifted feature in the real part for $q\approx1.25q_\textnormal{F}$ and $q\approx2.94q_\textnormal{F}$. 
Furthermore, we note that the imaginary part closely resembles the dynamic structure factor $S(q,\omega)$ [which is a direct consequence of the fluctuation--dissipation theorem, Eq.~(\ref{eq:FDT})], and the associated physics, thus, need not be further discussed at this point.

Let us next examine the difference between the three different depicted solutions for the dynamic density response function. Due to the strong electronic correlations, the RPA only provides a qualitative description, as it is expected. Including a local field correction does not only lead to a red-shift, but also to a significantly changed shape, and a substantially different static limit for Re$\chi$. For example, the mean field description of the dynamic density response predicts a shallow minimum in the real part around $\omega=1.5\omega_\textnormal{p}$ for $q\approx1.25q_\textnormal{F}$ (panel b), which is not present for both the static and the dynamic LFC. Finally, we note that the static LFC leads to a clear improvement over RPA in particular in the description of the peak position, but--as in the case of $S(q,\omega)$--cannot capture the nontrivial behaviour of $\chi(q,\omega)$ at $q\approx2.94q_\textnormal{F}$.

The bottom half of Fig.~\ref{fig:chi-rs10} corresponds to $r_s=4$ and $\theta=1$, which is located in the WDM regime. As for $S(q,\omega)$, both the real and imaginary part of $\chi(q,\omega)$ are not as sharply peaked as for $r_s=10$, as it is expected. Overall, the RPA seems to provide a somewhat better description of Re$\chi(q,\omega)$ than of Im$\chi(q,\omega)$, although there are substantial deviations for $\omega\to0$. Moreover, the static approximation is highly accurate and cannot be distinguished from the full solution for all three wave numbers.

\subsection{Dynamic dielectric function}\label{ss:results-epsilon}
Proceeding in a similar way as for the dynamic structure factor and the density response function, we now turn to a reconstruction of the dynamic dielectric function $\epsilon(q,\omega)$. This function is particularly interesting, as it gives direct access to the spectrum of collective excitations of the plasma. Using Eq.~(\ref{eq:eps-G}), the dynamic dielectric function is directly expressed by the local field correction to which we have access in our \textit{ab initio} simulations. Thus, it is straightforward to directly compare the RPA dielectric function to correlated  results that use either the static or dynamic LFC.

A first typical result for the dielectric function of the correlated electron gas is shown in Fig.~\ref{fig:eps-rs6}, for the case of $r_s=6$ and $\theta=1$. In the left (right) panel we show the real (imaginary) part of the dielectric function for two wave numbers. At large frequencies, $\omega \gtrsim \omega_p$ the correlated results are in close agreement with the RPA. However, strong deviations occur below $\omega_p$. The peak of the imaginary part narrows and shifts to much lower frequencies. Due to the Kramers-Kronig relations, the same trend is observed for the real part. The statistical uncertainty of the reconstruction of $G(q,\omega)$ leads to an uncertainty in the region of the peak of Im$\,\epsilon$ that is indicated by the red band. Interestingly, the static approximation is very close to the full dynamic results at the present parameters. In the left part of the figure, we also show the imaginary part of the inverse dielectric function, -Im$\,\epsilon^{-1}$ which is proportional to the dynamic structure factor, cf. Eq.~(\ref{eq:FDT}). At the lower wave number (top left figure), its peak is close to the zero of the real part of $\epsilon$.

\begin{figure}
    \centering
    \includegraphics{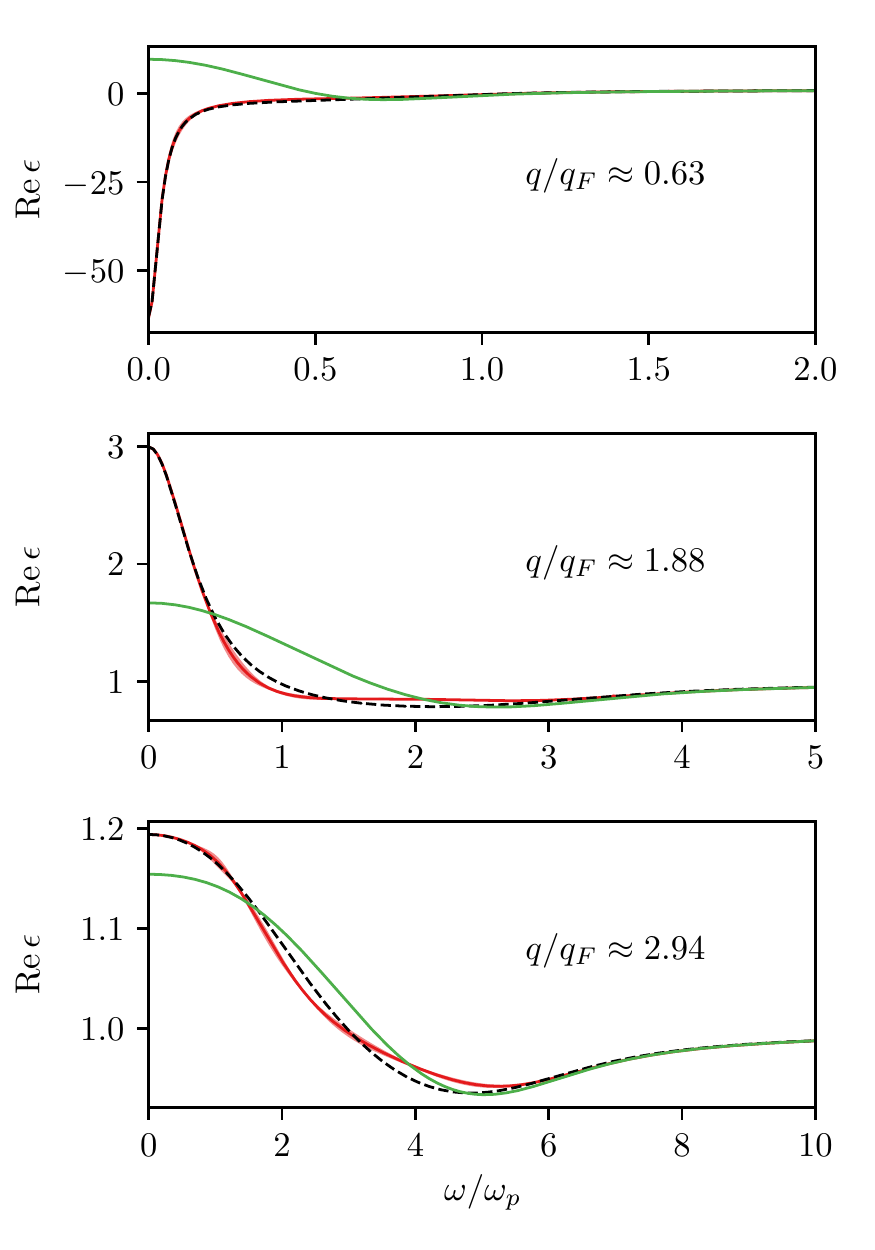}
    \caption{Real part of the dielectric function at different values of $q$ for $r_s=10$, $\theta=1$. Green lines: RPA, red (dashed) lines: dynamic (static) PIMC results. For small wave vectors, in the \textit{ab initio} results, the static limit becomes negative with a large absolute value, for details see text.}
    \label{fig:re-eps-rs10}
\end{figure}

\begin{figure}
    \centering
    \includegraphics{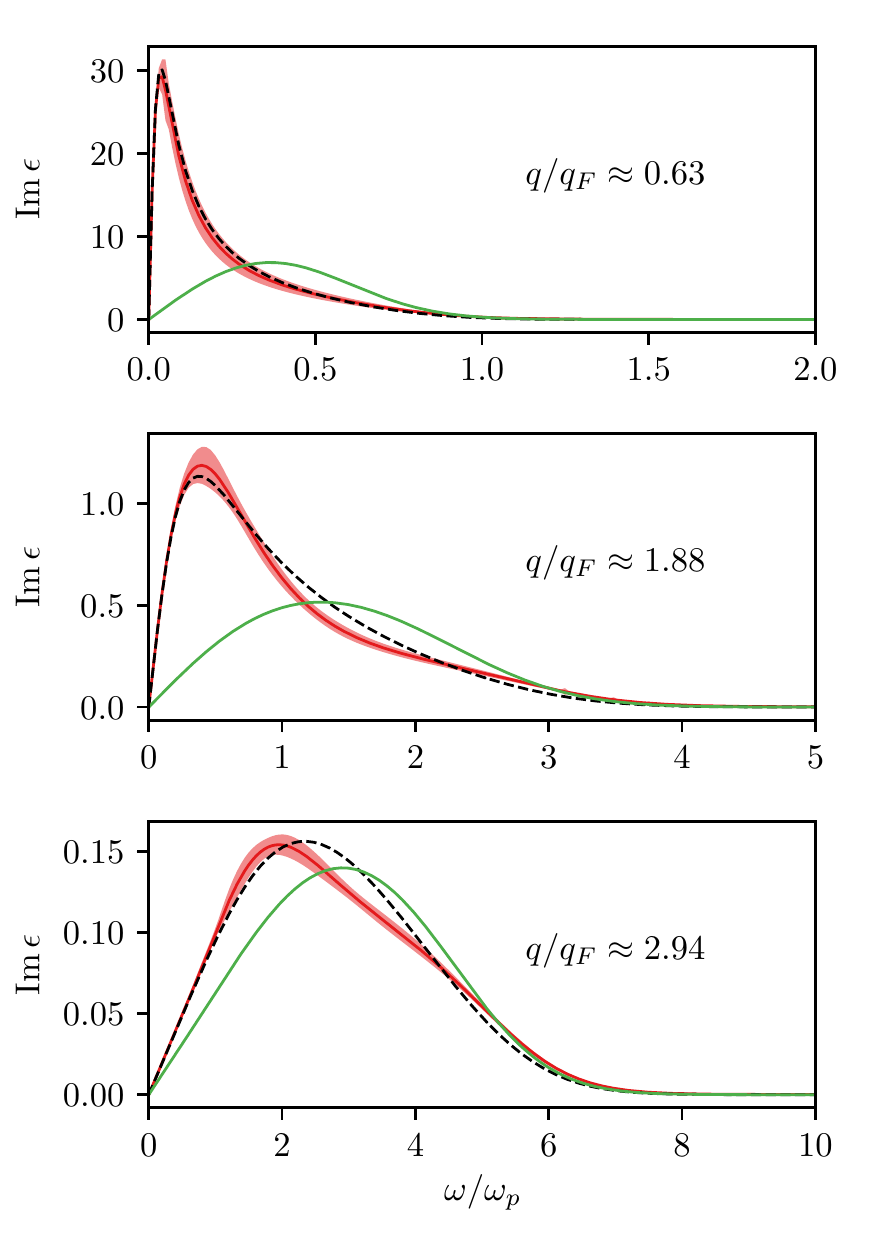}
    \caption{Imaginary part of the dielectric function for $r_s=10$, $\theta=1$. Green lines: RPA, red (dashed) lines: dynamic (static) PIMC results.
}
    \label{fig:im-eps-rs10}
\end{figure}

Let us now proceed to stronger coupling,  to the margins of the electron liquid regime ($r_s=10$ and $\theta=1$), and discuss the behavior of the dynamic dielectric function. Of particular interest is the possibility of  instabilities~\cite{quantum_theory}. 
 In Fig.~\ref{fig:re-eps-rs10}, we show the frequency dependence of Re$\,\epsilon(q,\omega)$, for $q\approx0.63q_\textnormal{F}$ (top), $q\approx1.88q_\textnormal{F}$ (center), and $q\approx2.94q_\textnormal{F}$ (bottom). For the two larger wave numbers, we find similar trends as for $r_s=6$, although the differences between the RPA and the LFC based curves is substantially larger, in particular around $q=2q_\textnormal{F}$. The top panel, on the other hand, exhibits a peculiar behavior, which deserves special attention: while the RPA predicts a positive static limit, as it is expected, the red and black curves attain a comparably large (though finite) negative value, for $\omega=0$.

\begin{figure}[h]
    \centering
    \includegraphics[width=0.9\columnwidth]{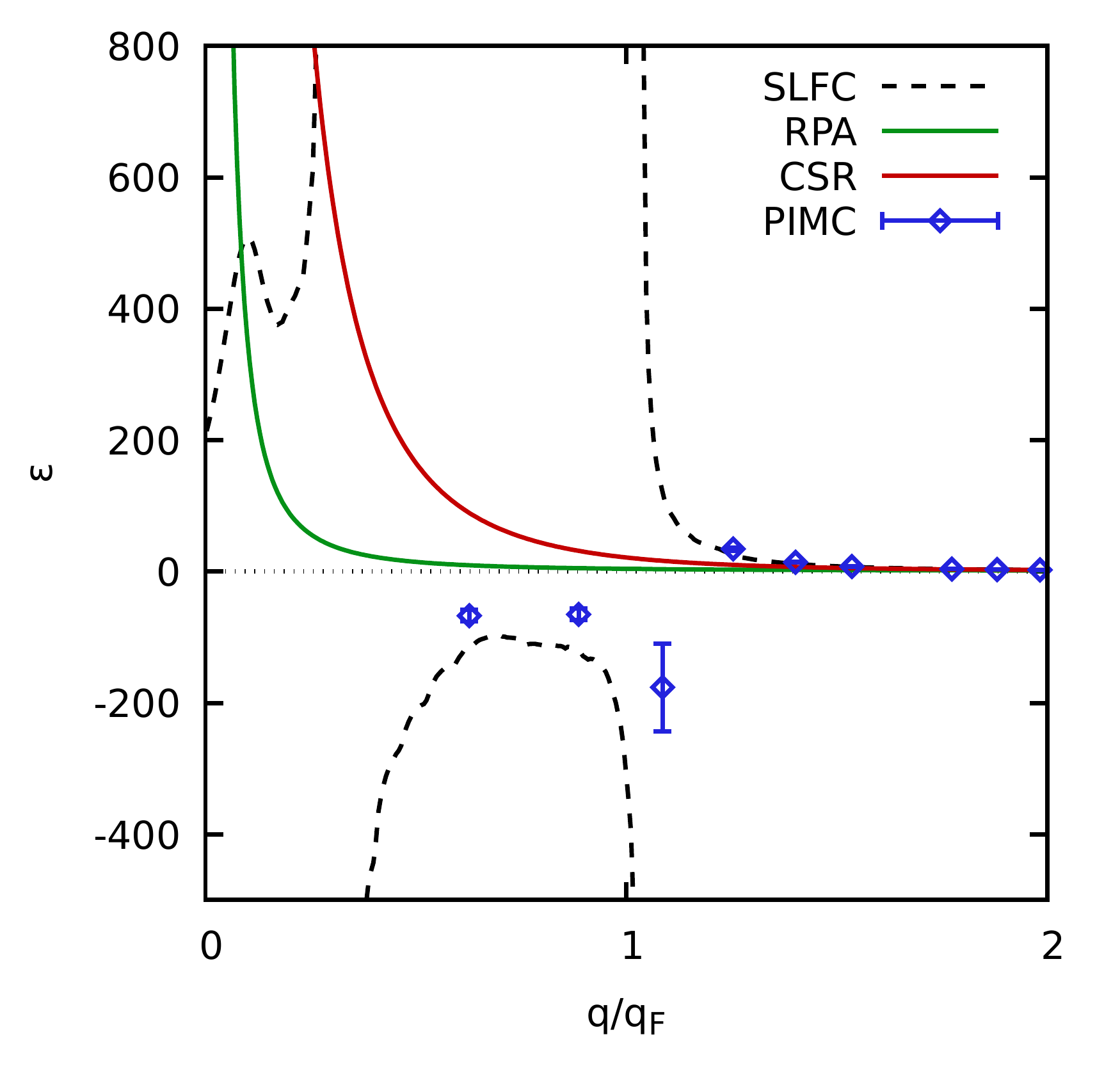}
    \caption{Static limit of the dielectric function at $r_s=10$ and $\theta=1$. The blue diamonds depict PIMC data for $N=34$ electrons and the solid green and dashed black curves the RPA results and the neural-net representation of the static LFC from Ref.~\cite{dornheim_ML}. Moreover, the solid red curve has been obtained by using for $G(q)$ the compressibility sum-rule given by Eq.~(\ref{eq:real_CSR}).}
    \label{fig:static_epsilon_rs10_TD}
\end{figure}

To understand the implications of this nontrivial finding, we show the wave number dependence of the static limit of Re$\chi(q,\omega)$ in Fig.~\ref{fig:static_epsilon_rs10_TD}. The  RPA curve (green line) converges to $1$ from above, for large $q$ and diverges to positive infinity for $q\to0$, as it is known from theory~\cite{quantum_theory}. The static approximation (dashed black), on the other hand, leads to an altogether different behaviour. While it eventually attains the same limit for large $q$-values, it exhibits a highly nontrivial structure with two poles around $q\approx q_\textnormal{F}$ and $q\approx0.4q_\textnormal{F}$, and remains finite for $q=0$ (with a small local maximum around $q\approx0.1q_\textnormal{F}$). Let us first briefly touch upon the implications of this behaviour for the stability of the system. As we have noted in Sec.~\ref{ss:epsilon}, the static dielectric function needs to remain outside the interval between zero and one, $\epsilon(q,0) \not\in (0,1]$. This requirement is indeed fulfilled by the static approximation, since the sign changes are around singular points. Yet, the finite value for $q\to0$ clearly violates the exact constraint given in Eq.~(\ref{eq:compressibility}) above and deserves further attention. In particular, Eq.~(\ref{eq:compressibility}) can be re-phrased in terms of the LFC as
\begin{eqnarray}
\lim_{q\to0}\epsilon(q,0) &=& \lim_{q\to0} 1 - \frac{\frac{4\pi}{q^2}\chi_0(q,0)}{1+\frac{4\pi}{q^2}G(q,0)\chi_0(q,0)}\\ \nonumber
&=& \lim_{q\to0} 1 -
\frac{4\pi\chi_0(q,0)}{q^2[1+4\pi C \chi_0(q,0)]}\ ,
\end{eqnarray}
where the second equality follows from the well-known limit of the static local field correction
\begin{eqnarray}
\lim_{q\to0}G(q,0) = q^2 C\ , \label{eq:real_CSR}
\end{eqnarray}
see, e.g., Refs.~\cite{dynamic_folgepaper,dornheim_ML} for details. Naturally, this limit was incorporated into the training procedure for the neural-network representation of $G(q,0)$ in Ref.~\cite{dornheim_ML}.
However, directly utilizing Eq.~(\ref{eq:real_CSR}) to compute a dielectric function (which becomes exact for $q\to0$) leads to the red curve in Fig.~\ref{fig:static_epsilon_rs10_TD}, which does indeed diverge towards positive infinity as predicted by Eq.~(\ref{eq:compressibility}). The explanation for the finite value at $q=0$ (and also the unsmooth behaviour for small $q$) of the dashed black curve is given by the construction of the neural-network representation itself. As any deviations both from Eq.~(\ref{eq:real_CSR}) for small $q$, and the \textit{ab initio} PIMC input data elsewhere, were equally ``punished'' by the loss function, the resulting neural network exhibits an overall absolute accuracy of $\Delta G\sim0.01$ and, thus, does not exactly go to zero in the long wave-number limit. 
We, thus, conclude that using the neural-network representation from Ref.~\cite{dornheim} leads to the exact $q\to0$ limit for density response quantities like $\chi(q)$ and $S(q)$, but becomes inaccurate for effective, dielectric properties like $\epsilon(q)$ and $\Pi(q)$ in this regime.

\begin{figure}
    \centering
    \includegraphics[width=0.9\columnwidth]{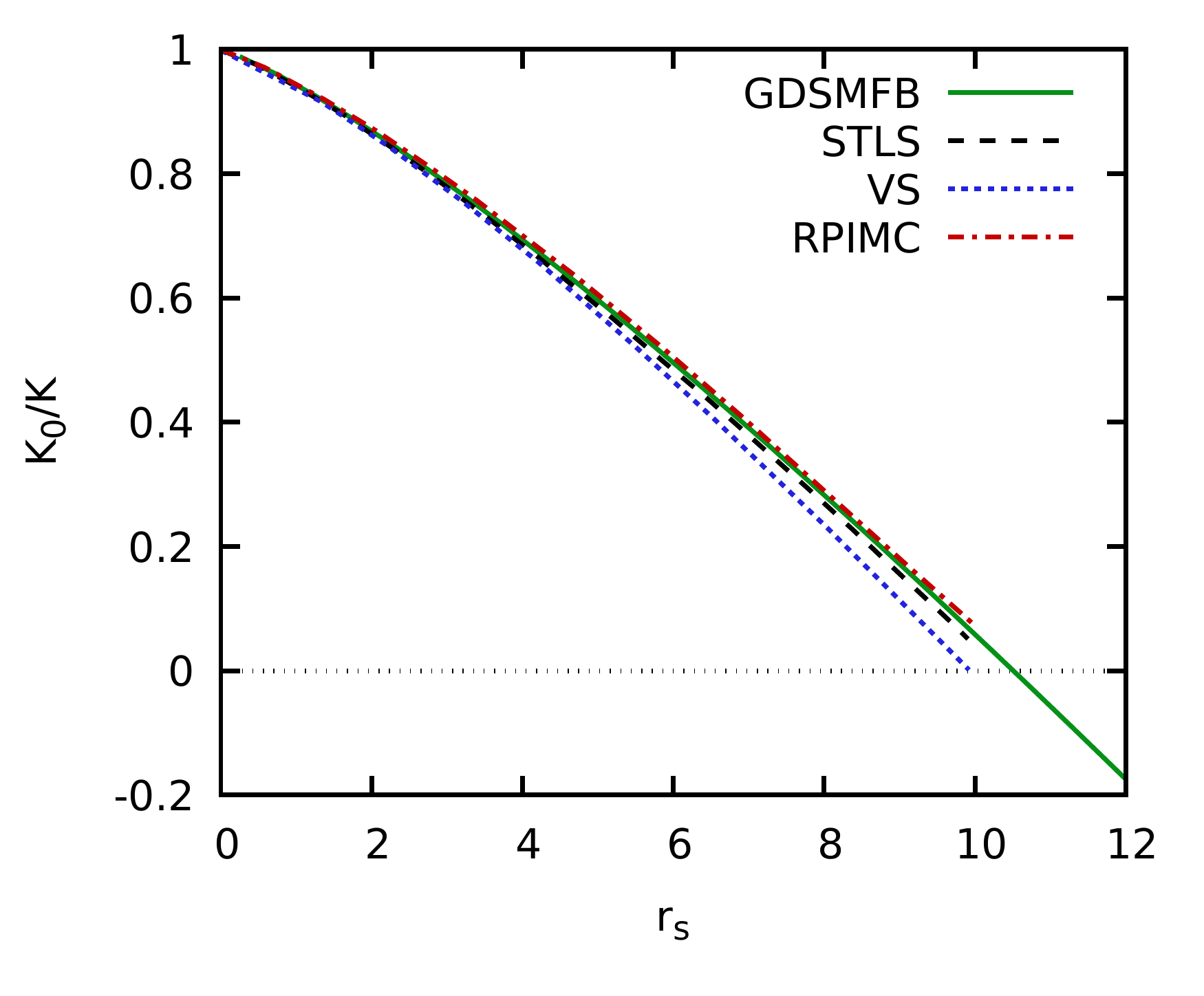}
    \caption{Density dependence of the ratio of the ideal to the interacting compressibility for $\theta=1$. The solid green line has been obtained from the parametrization of $f_\textnormal{xc}$ by Groth \textit{et al.}~\cite{groth_prl}, and the dashed black, dotted blue, and dash-dotted red lines have been taken from Ref.~\cite{stls2} and correspond to STLS, VS, and restricted PIMC, respectively.}
    \label{fig:CSR}
\end{figure}

Let us conclude this section with a brief discussion of the eponymous quantity of Eq.~(\ref{eq:compressibility}), i.e., the compressibility $K$. In Fig.~\ref{fig:CSR}, we show the $r_s$-dependence of the ratio of the noninteracting to the interacting compressibility, for $\theta=1$, and the solid green line has been obtained from the accurate PIMC-based parametrization of $f_\textnormal{xc}$ by Groth and co-workers~\cite{groth_prl}. In addition, we also show results from (static) dielectric theories investigated by Sjostrom and Dufty~\cite{stls2}, namely STLS~\cite{stls,stls_original} (dashed black) and VS~\cite{vs_original,stolzmann} (dotted blue). Finally, the dash-dotted red curve has been obtained by the same authors on the basis of the restricted PIMC data by Brown \textit{et al.}~\cite{brown_ethan}.

First and foremost, we note that all four curves exhibit a qualitatively similar trend and approach the correct limit for $r_s\to0$, where the UEG becomes ideal. With increasing coupling strength, the compressibility is reduced as compared to $K_0$. Moreover, $K$ does eventually become negative around $r_s\approx10.5$. This has some interesting implications and indicates, that the static dielectric function $\epsilon(q,0)$ converges towards negative infinity in the long wavelength limit. Lastly, we compare the three approximate curves from Ref.~\cite{stls2} against the accurate GDSMFB-benchmark and find that RPIMC and STLS exhibit relatively small systematic deviations, whereas the VS-curve deviates the most. This is certainly remarkable as the closure relation of the VS-formalism strongly depends on a consistency relation of $K$ towards both $G(q,0)$ and $f_\textnormal{xc}$, see Ref.~\cite{stls2} for a detailed discussion of this point.

\begin{figure*}
\centering
\includegraphics[width=0.9\linewidth]{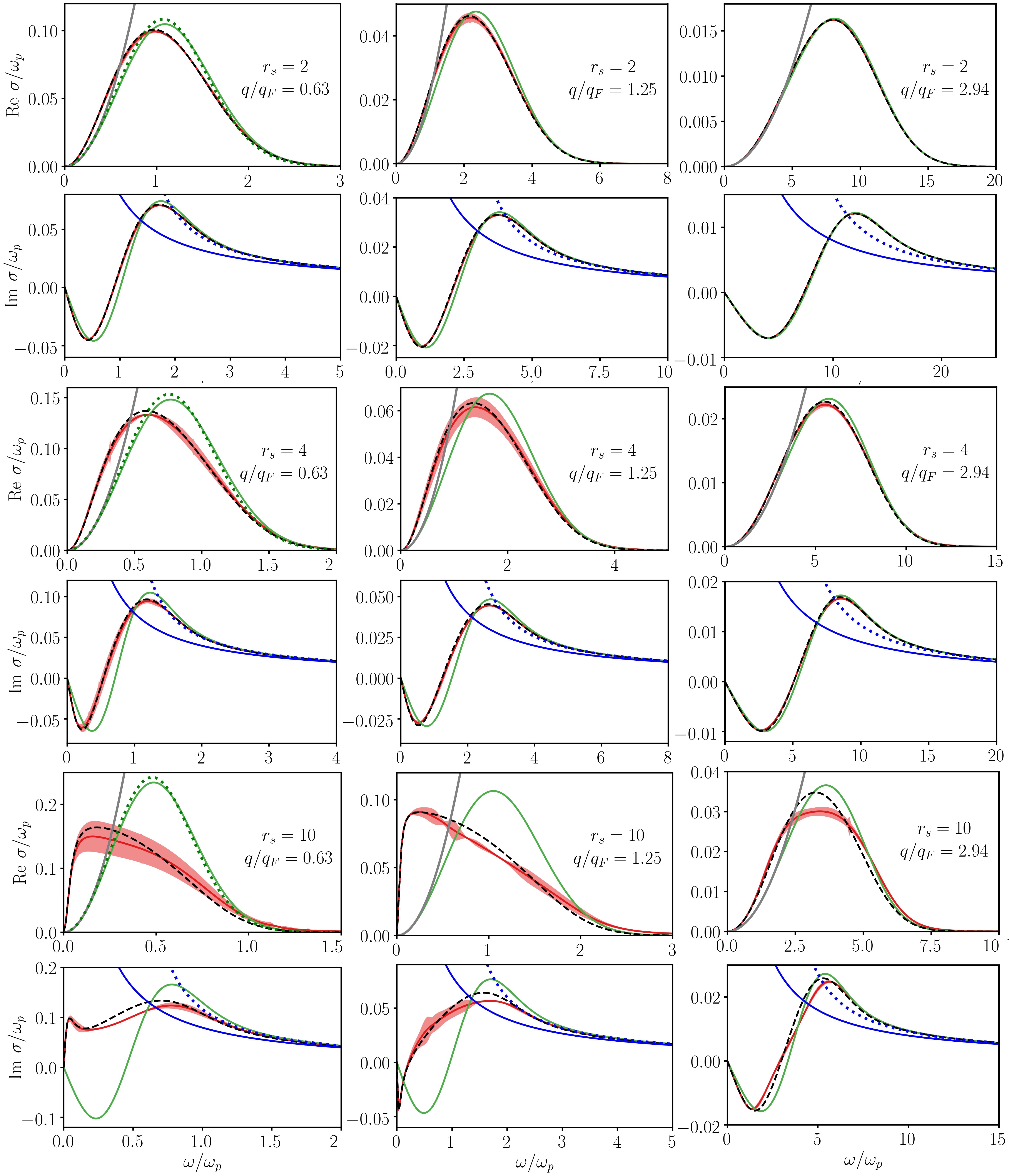}
\caption{Real and imaginary part of the dynamic conductivity at $\theta=1$ for three densities -- Top: $r_s=2$, middle: $r_s=4$, bottom: $r_s=10$ -- and three wavenumbers -- left: $q/q_F\approx0.63$, middle $q/q_F\approx1.25$, and right: $q/q_F\approx2.94$.
\textit{Ab initio} simulation results using the static (black dashed line) and dynamic (red bands) LFC are compared to the  RPA (full green line) and analytical asymptotics:  For Re $\sigma$ approximations Eq.~(\ref{eq:RPA_limit1}) and  Eq.~(\ref{eq:RPA_limit2}) are shown by the solid gray and dotted green line (for $q/q_F\approx0.63$ only), respectively. For Im $\sigma$ the approximations  Eq.~(\ref{eq:RPA_limit3}) 
and Eq.~(\ref{eq:RPA_limit4}) are shown by the dotted blue and solid blue lines, respectively.   }
\label{fig:conductivity}
\end{figure*}

\subsection{\textit{Ab initio} results for the dynamic conductivity}\label{ss:other}

Let us start our investigation of the dynamic conductivity with an analysis of the four asymptotics introduced in Sec.~\ref{ss:conductivity}. This is shown in Fig.~\ref{fig:conductivity} corresponding to $r_s = 2, 4$, and $r_s=10$ and three wavenumbers. The agreement to the full RPA solution (solid green curve) is very good. First, we note that it is evident from Eq.~(\ref{eq:RPA_limit2}) that the real part of the RPA conductivity tends to zero, if the limit  $q\to0$ is taken (optical conductivity), cf. dotted green lines in the plots for $q/q_F\approx0.63$. 
Second, Re $\sigma^{\rm RPA}$
 tends to zero as $\omega^2$ if the limit $\omega\to0$ is taken for arbitrary $q$ (grey lines), cf. Eqs.~(\ref{eq:RPA_limit2}) and (\ref{eq:RPA_limit1}).  
 Third, the asymptotic behaviour from Eq.~(\ref{eq:RPA_limit3}) (cf. dotted blue lines) correctly reproduces the behavior of Im $\sigma^{\rm RPA}$ for frequencies larger than the first positive maximum, for all $q$, and with further frequency increase merges with   Eq.~(\ref{eq:RPA_limit4}) (solid blue line) which correctly captures the high frequency limit.   

Having \textit{ab initio} results for the local field correction and the longitudinal polarization available allows us to produce accurate results for the conductivity including exchange and correlation effects as well.
Simulation results for the conductivity within the static and dynamic LFC approximation are presented in Fig.~\ref{fig:conductivity} by the black dashes lines and red bands, respectively.
First, we observe that the static and dynamic results for the conductivity are in very good agreement with each other for all cases. Small deviations are visible mainly for $r_s=10$. Second, the agreement of the RPA conductivity with the correlated approximations is reasonable for $r_s=2$. At stronger coupling, $r_s=4$, good agreement is observed only at the largest wavenumber, $q/q_F\approx2.94$, whereas for lower $q$ deviations are growing. At $r_s=10$ agreement is observed only at large frequencies, whereas the behavior around the main peak of Re $\sigma$ as well as for frequencies below the peak shows dramatic influences of correlations, except for the largest $q$.

We note that the present analysis is complementary to the approaches for the conductivity executed in the optical limit, $q\to 0$. In this limit, the focus is usually on incorporating electron-ion correlations that can be shown to be the dominant non-ideality contribution in that  limit. Examples include linear response calculations as well as DFT approaches using Kubo-Greenwood relations~\cite{Reinholz_2000,Veysman:2016,Desjarlais:2017,low_density1,low_density2,holst_2008}.
Electron-electron and electron-ion correlations have been taken into account simultaneously, but only for the non-degenerate, weakly coupled case \cite{Desjarlais:2017}. The current work opens up the possibility to improve this description and extend it to the warm dense matter regime with all its correlations and quantum effects causing a non-Drude form of the conductivity.

\section{Summary and discussion\label{sec:summary}}

In this work, we have investigated in detail the calculation of dynamic properties based on \textit{ab initio} PIMC data for the warm dense electron gas. More specifically, we have discussed the imaginary-time version of the intermediate scattering function $F(q,\tau)$, which is used as a starting point for the reconstruction of the dynamic structure factor $S(q,\omega)$. This is achieved by a recent stochastic sampling scheme of the dynamic local field correction~\cite{dornheim_dynamic,dynamic_folgepaper}, which, in principle, gives the complete wave-number- and frequency- resolved description of exchange--correlation effects in the system.

In particular, we have demonstrated that such knowledge of $G(q,\omega)$ allows for the subsequent accurate calculation of other dynamic quantities such as the dynamic density response function, $\chi(q,\omega)$, the (inverse) dielectric function $\epsilon(q,\omega)$ [$1/\epsilon(q,\omega)$], and the dynamic conductivity $\sigma(q,\omega)$.
Therefore, our new results will open up avenues for future research, as they contain key information about different properties of the system: $\chi(q,\omega)$ fully describes the response of the system to an external perturbation, e.g. by a laser beam~\cite{quantum_theory,dornheim_nonlinear}; the dielectric function is of paramount importance in electrodynamics and gives access to the full spectrum of collective excitations~\cite{bonitz_book}. This is a fundamental point, which will be explored in detail in a future publication~\cite{hamann_df_20}; the dynamic conductivity in warm dense matter is of particular importance for magneto-hydrodynamics, e.g., planetary dynamos~\cite{wicht_2018}.

An additional key point of this paper is the investigation of the so-called \emph{static approximation}~\cite{dornheim_dynamic,dynamic_folgepaper}, where the the exact dynamic LFC is replaced by its exact static limit $G(q,0)$ that has recently become available as a neural-net representation~\cite{dornheim_ML}. Here we found that this approach allows basically for exact results for all dynamic quantities mentioned above over the entire $q$- and $\omega$-range for $r_s\lesssim4$, i.e., over substantial parts of the WDM regime. This has important applications for many aspects of WDM theory such as the on-the-fly interpretation of XRTS experiments~\cite{siegfried_review,kraus_xrts}, as it comes with no additional computational cost compared to RPA. For larger values of $r_s$, the static approximation does induce significant deviations to the exact results, but it nevertheless reproduces the most important trends of the various dynamical properties that are absent in an RPA-based description.

Our investigation of the dynamic dielectric function  has uncovered that electronic exchange--correlation effects [either by using $G(q,\omega)$ or $G(q,0)$] lead to a nontrivial behavior of $\epsilon(q,\omega)$, where the static limit can actually become negative for certain wave numbers. This is i) a correct physical behaviour and ii) does not signal the onset of an instability, and neither does the negative compressibility depicted in Fig.~\ref{fig:CSR}. In contrast, our analysis of the full wave-number dependence of $\epsilon(q,0)$ has revealed that the finite accuracy of the neural-net representation~\cite{dornheim_ML} of $G(q,0)$ does induce an artificial, unphysical behaviour in the $q\to0$ limit of this quantity. Yet, this does not constitute a fundamental obstacle and could potentially be removed by replacing the neural net with the exact compressibility sum-rule in this regime. Moreover, we mention that directly observable quantities like $S(q,\omega)$, $S(q)$, or $\chi(q)$ are not afflicted with this issue, as here the impact of the LFC vanished for small wave numbers.
Lastly, the conductivity is afflicted the same way as the dielectric function, but this might be alleviated if e-i scattering is included~\cite{Reinholz_2000}.

Future extensions of our research include the implementation of other imaginary-time correlation functions into our PIMC simulations. Possible examples are given by the Matsubara Green function~\cite{boninsegni1,dynamic_alex1} or the velocity autocorrelation function~\cite{velocity}, which would give access to the single-particle spectrum or a dynamical diffusion constant, respectively. Furthermore, the combination of PIMC simulations with the reconstruction scheme explored in this work can potentially be applied to real electron-ion-plasmas, which would allow for the first time to compute \textit{ab initio} results of, e.g., XRTS signals that can be directly compared to state-of-the-art WDM experiments.

\section*{Acknowledgments}

This work is supported by the German Science Foundation (DFG) via grant BO1366-13.
T.~Dornheim acknowledges support by the Center of Advanced Systems Understanding (CASUS) which is financed by Germany’s Federal Ministry of Education and Research (BMBF) and by the Saxon Ministry for Science, Culture and Tourism (SMWK) with tax funds on the basis of the budget approved by the Saxon State Parliament. 
Zh.A.~Moldabekov  acknowledges support via the Grant  AP08052503 by the Ministry of Education and Science of the Republic of Kazakhstan. 

All PIMC calculations were carried out on the clusters \emph{hypnos} and \emph{hemera} at Helmholtz-Zentrum Dresden-Rossendorf (HZDR), the computing centre of Kiel university, at the Norddeutscher Verbund f\"ur Hoch- und H\"ochstleistungsrechnen (HLRN) under grant shp00015, and on a Bull Cluster at the Center for Information Services and High Performance Computing (ZIH) at Technische Universit\"at Dresden.

\section*{Appendix}\label{s:appendix}
Here we present approximate results for the second and fourth moments of the finite temperature Fermi function. For the second moment, one finds
\begin{equation}
    \frac{\langle v^2\rangle}{v_F^2}=\frac{3}{2} \theta^{5/2} I_{3/2}(\eta)\simeq \frac{3}{2}\frac{\left(\frac{4}{25}+\theta^2\right)^{1/2}}{1-0.14\left[e^{-\theta}-e^{-3.68 ~\theta}\right]},
    \label{eq:1_moment}
\end{equation}
where $I_\nu$ is the Fermi integral of order $\nu$. Similarly, for the fourth moment follows
\begin{equation}
    \frac{\langle v^4\rangle}{v_F^4}= \frac{3}{2} \theta^{7/2}I_{5/2}(\eta)\simeq \frac{3}{7}\left(1+6.4875~\theta^{\frac{2}{1.16}}\right)^{1.16}.
     \label{eq:2_moment}
\end{equation}
The parametrizations in Eqs.~(\ref{eq:1_moment}) and (\ref{eq:2_moment})
agree with the exact numerical results with a precision better than $3\%$, in the entire range of $\theta$.

\end{document}